\documentclass[draft]{agujournal2019}
\usepackage{url} 
\usepackage{lineno}
\usepackage{soul}
\usepackage{hyperref}
\usepackage{natbib}
\usepackage{amsmath}
\usepackage{amssymb}
\usepackage{pdflscape}

\newcommand{\newTxt}{\textcolor{black}}
\renewcommand{\st}[1]{\unskip}

\newcommand{\bs}{\boldsymbol{s}}
\newcommand{\pkg}[1]{{\fontseries{b}\selectfont #1}}
\let\proglang=\textsf

\draftfalse
\journalname{ArXiv.org}

\begin{document}

%
%


\title{Data-driven upper bounds and event attribution for unprecedented heatwaves}

%
%

\authors{Mark D. Risser\affil{1}, Likun Zhang\affil{2}, Michael F. Wehner\affil{3}}

\affiliation{1}{Climate and Ecosystem Sciences Division, Lawrence Berkeley National Laboratory, Berkeley, CA, 94720.}
\affiliation{2}{Department of Statistics, University of Missouri, Columbia, MO, 65211.} 
\affiliation{3}{Applied Mathematics and Computational Research Division, Lawrence Berkeley National Laboratory, Berkeley, CA, 94720.}

\correspondingauthor{Mark D. Risser}{mdrisser@lbl.gov}

\begin{keypoints}
\item 70\% of the most extreme heatwaves are deemed ``impossible'' due to \newTxt{use of suboptimal} \st{poor} analysis methods.

\item Improved methods reveal 21 historical heatwaves are ``impossible'' without climate change. 

\item Inland heatwaves can include temperatures that are hotter than 7$\sigma$ events. 

\end{keypoints}


\begin{abstract} 
The last decade has seen numerous record-shattering heatwaves in all corners of the globe. In the aftermath of these devastating events, there is interest in identifying worst-case thresholds or upper bounds that quantify just how hot temperatures can become. 
Generalized Extreme Value theory provides a data-driven estimate of extreme thresholds; however, upper bounds may be exceeded by future events, which undermines attribution and planning for heatwave impacts.
Here, we show how the occurrence and relative probability of observed yet unprecedented events that exceed \textit{a priori} upper bound estimates, so-called ``impossible'' temperatures, has changed over time.
We find that many unprecedented events are actually within data-driven upper bounds, but only when using modern spatial statistical methods. 
Furthermore, there are clear connections between anthropogenic forcing and the ``impossibility'' of the most extreme temperatures. 
Robust understanding of heatwave thresholds provides critical information about future record-breaking events and how their extremity relates to historical measurements. 
\end{abstract}

\section{Introduction} \label{sec:intro}

In recent years, our planet has experienced a growing number of record-breaking heatwaves that have a devastating impact on human health and infrastructure. Western Russia experienced temperatures in June, 2010, that were unprecedented since at least the 19th century \citep{Rahmstorf2011}, which contributed to significant loss of human life \citep{Dole2011}. A deadly heatwave impacted much of western Europe in June and July, 2019, breaking previous records in metropolitan France by nearly 2$^\circ$C and directly causing hundreds of excess deaths \citep{Mitchell2018extreme, Vautard2020}. In late June, 2021, an unprecedented heatwave struck the United States Pacific Northwest and western Canada that broke all-time records by more than 15$^\circ$C  \citep{emily2022dynamics} and was among the most extreme events ever recorded globally \citep{Thompson2022western}, causing over 500 deaths \citep{popovich2021hidden} and significant agricultural losses \citep{Baker21}. These and many other similarly devastating heatwaves are termed ``low likelihood, high-impact'' (LLHI) weather events by the sixth Intergovernmental Panel on Climate Change Report, which furthermore states that we currently have low confidence in current and future projections of LLHIs \citep{Seneviratne2022_book}. It is now clear that large swaths of the globe are vulnerable to LLHI heatwaves \citep{Thompson2023most}, which poses serious problems for adaptation and impacts planning.

In the aftermath of a devastating heatwave event, there is considerable interest in quantifying worst-case upper bounds on extreme hot temperatures as well as their frequency or return interval. Atmospheric theory provides physical upper bounds on extreme temperatures \citep[see, e.g.,][]{Zhang2023upper}, but these generally correspond to idealized conditions that may rarely occur in reality. Generalized extreme value (GEV) theory provides a data-driven estimate of thresholds for extremes: when the GEV shape parameter is negative, the distribution has a finite upper bound \citep{Coles2001}. When the shape parameter is less than $-0.3$, the distribution is very sharply bounded. However, for the aforementioned heatwaves in Russia, France, and British Columbia, the hottest temperatures experienced during the event exceeded \textit{a priori} GEV-based upper bounds. This leads us to define a so-called ``impossible'' temperature: measurements that are extreme enough relative to more typical extrema that they exceed what was previously thought to be the hottest possible temperatures. We use ``impossible'' informally since, of course, the temperatures actually occurred in Nature. Impossible temperatures defined this way pose serious problems to planning for the impacts of extreme heatwaves: the best estimate is that these events have a zero probability, meaning we cannot assess their rarity or return interval much less determine the extent to which human activities affected the statistics of the event. 

In many cases, the data-driven upper bounds used to qualify a temperature as ``impossible'' are derived using traditional extreme value methods: 
non-stationary GEV analysis with a monotonically increasing covariate such as time or greenhouse gas concentrations to account for climate change, applied independently to the records at each weather station
\citep[see, e.g,][]{van2019human,Philip21,emily2022dynamics}. It has been shown that augmenting time trends with additional physically-based covariates can often help anticipate the most extreme temperatures \citep{Zeder2023quantifying}. While the traditional approach (even when augmented with additional covariates) is relatively straightforward to implement and hence broadly used, it ignores an obvious source of information: measurements of temperature extremes from nearby locations. One solution to address this limitation leverages a relatively old idea, wherein one ``trades space for time'' as is done in regional frequency analysis \citep{Hosking1993}. This is particularly important given that the relatively short observational record can lead to real-world events being deemed impossible \citep{Zeder2023}. The broad statistical literature on novel extreme value techniques \citep[see, e.g.,][]{huser2019modeling, zhang2021hierarchical, zhang2023flexible} allows us to use information from spatially-nearby sites and provides a path forward for assigning non-zero probabilities to the most extreme temperatures, even from a data-driven perspective.

In this paper, we explore how methodological choices impact data-driven upper bound thresholds for the most extreme temperatures\newTxt{, what we refer to as \textit{unprecedented} heatwaves.} We specifically focus on one-day temperature extremes, since their underlying analyses commonly form the basis for climate change adaptation, mitigation, and attribution. \newTxt{``Unprecedented'' is defined as annual maximum daily maximum temperatures that are at least $4\sigma$ events relative to other annual maxima (see Section~\ref{sec:selectImposs}).} Our approach leverages in situ records and is distinct from dynamical model-based studies to assess statistics of impossible temperatures \citep[e.g.,][]{McKinnon2022,Fischer2023}. We explicitly demonstrate that, using the same input data, it is \textit{very likely} that state-of-the-art methods can explain 69.5\% more of the impossible temperatures from the historical record relative to the traditional approach. Equipped with quantitatively non-zero probabilities for these events, we then revisit the attribution problem and propose a metric to quantify the relative rarity of the next unprecedented heatwave event.

\section{Materials and Methods}

\subsection{Data sources} \label{sec:covariates}
\st{We analyze measurements of daily maximum temperature ($^\circ$C) from the Global Historical Climate Network-Daily (GHCN-D) database (Menne et al., 2012)
over the historical record, defined as 1901 to 2022. We identify a high-quality set of $N=7,992$ records based on a minimum threshold of non-missing daily measurements (see Supporting Information Section 1.1 for more details). The geographic distribution of these stations is shown in Supplemental Figure S1. We then obtain an extreme value sample from the annual maximum daily maximum temperature measurement (denoted ``TXx'') from each station. Across all stations and years, this yields $n = 612,735$ TXx measurements for analysis.}

\newTxt{We analyze measurements of daily maximum temperature ($^\circ$C) from the Global Historical Climate Network-Daily (GHCN-D) database \citep{Menne2012} over the historical record, defined as 1901 to 2022. We identify a high-quality set of records based on a minimum threshold of non-missing daily measurements as follows: first, we define annual ``blocks'' as January-December for stations in the Northern Hemisphere and July-June for stations in the Southern Hemisphere. Next, we calculate and store the maximum daily maximum temperature (denoted ``TXx'') in each block-year so long as that block-year had at least 66.7\% non-missing daily measurements. We also require that the TXx occurs in the warm season, i.e., April-September for the Northern Hemisphere and October-March for the Southern Hemisphere. We then select stations that have at least 50 years of non-missing TXx measurements over 1901-2022. Finally, we remove stations that have less than one station within approximately 500km, since this prevents us from leveraging spatially-nearby measurements (this step removes 121 stations from the 8,113 records with at least 50 years of data). Ultimately, this yields $N=7,992$ gauged locations for analysis, denoted $\mathcal{S}$, the geographic distribution of which is shown in Supplemental Figure S3. Denote the TXx measurements as $\{ Y(\bs,t) \}$ in year $t = 1901, \dots 2022$ for station ${\bf s} \in \mathcal{S}$. Across all stations and years, this yields $n = 612,735$ non-missing TXx measurements for analysis.}
\newTxt{It is important to note that the temperature records we analyze are of differing length: of the $7,992$ records, only around $2,000$ have non-missing TXx measurements in the early 1900s and only around $4,000$ in 2022; the highest density of non-missing records was in the 1960s (see Figure~\ref{fig_where_when}a.). The geospatial distribution of the overall length of temperature records and for each decade is shown in Supplemental Figure S4.}

Physical information about the Earth system is a useful tool for describing spatial and spatio-temporal variability in the behavior of weather extremes \citep[see, e.g.,][]{Zhang2010influence, Risser2021quantifying, Zeder2023quantifying, zhang2023explaining}.
First, we use five covariates to describe year-to-year variability and secular trends in the TXx climatology: a time series of the radiative forcing from greenhouse gases (GHGs) to describe human-induced secular trends \citep[following seminal work from][]{arrhenius1897influence}; the ENSO Longitude Index \citep[ELI,][]{Williams2018diversity} to account for the El Ni\~no-Southern Oscillation (ENSO); the Standardised Precipitation Evapotranspiration Index \citep[SPEI;][]{vicente2010multiscalar} to account for the effect of evapotranspirative cooling from the surface soil moisture content and local vegetation \citep{Domeisen2023}; and the Pacific-North American (PNA) teleconnection pattern and the North Atlantic Oscillation (NAO) to account for large-scale modes of climate variability \citep{Kenyon2008}. 
Second, we utilize elevation (meters above sea level) to describe orographically-driven heterogeneity in the statistical parameters that define the climatological distributions. 
For more information, see Section 1.2 of the Supporting Information and Supplemental Figure S5.

\subsection{Selecting a test set of unprecedented temperatures} \label{sec:selectImposs}

\begin{figure}[!t]
    \centering
    \includegraphics[width=\textwidth]{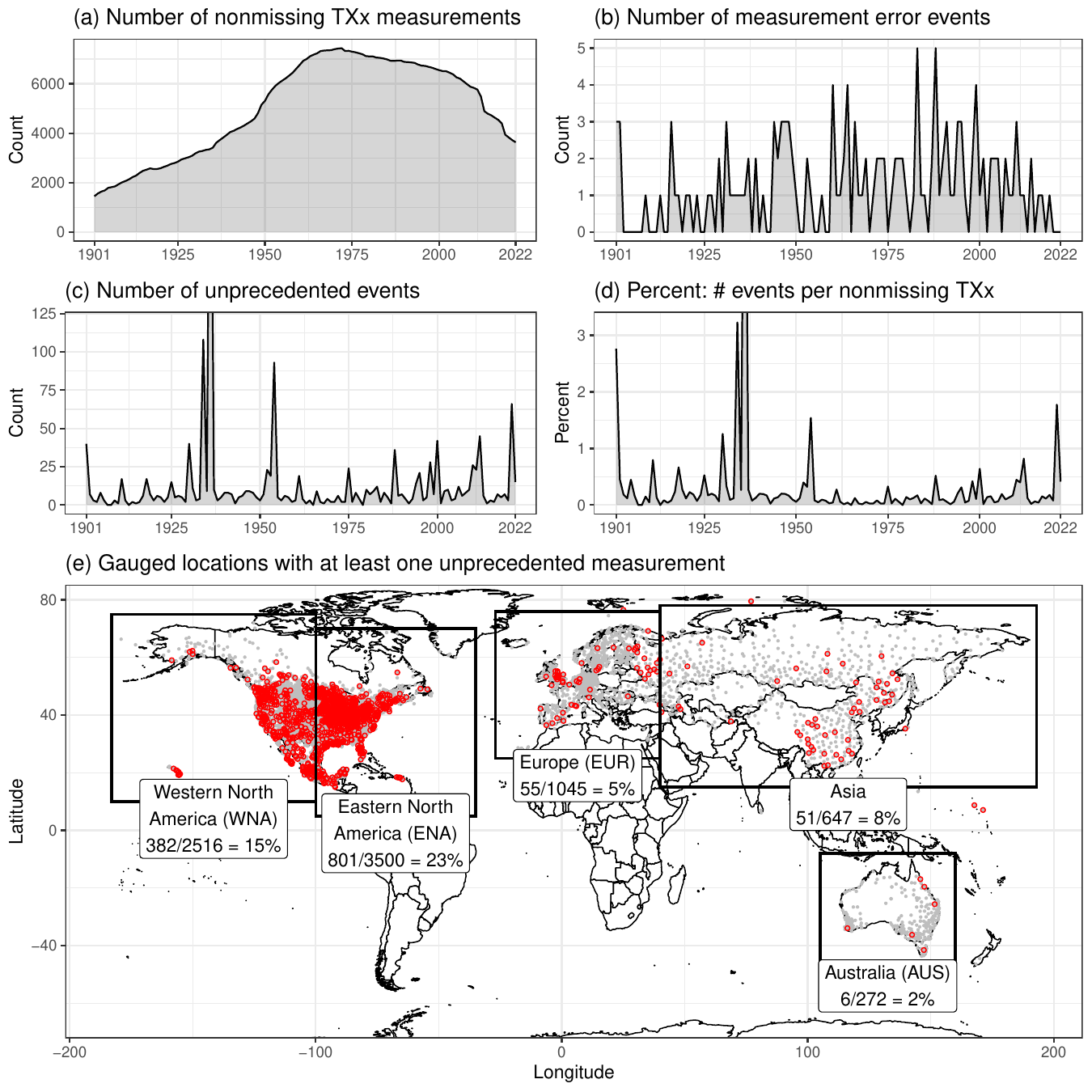}
    \caption{The number of ``real'' impossible measurements and measurement errors in each year (panels a. and b., respectively), along with the geographic distribution of GHCN-D gauged locations (gray points, all records) with at least one \st{real-impossible} \newTxt{unprecedented} event (red circles; panel c.). Boundaries for five continental summaries are also shown in panel (c) along with the fraction of stations in each region with at least one \st{real-impossible} \newTxt{unprecedented} event.}
    \label{fig_where_when}
\end{figure}

To assess the efficacy of data-driven upper bounds derived using statistical methods, we need a set of test events \newTxt{that are excluded from the analysis} -- those that \st{might be ``impossible,'' i.e.,} are \newTxt{very} extreme \newTxt{(and hence ``unprecedented'')} even relative to more typical annual maxima. 
\st{As in Section 1, we use ``impossible'' informally, since these events occurred in Nature.}
For our analysis, it is particularly important to avoid the selection bias associated with so-called ``trigger events'' \citep{Miralles2023}, which can lead to errors in return level estimates of the most extreme events. 
In selecting a set of test events from gauge-based records, however, it is critical to ensure that the selected events are \st{``}real\st{''} and do not correspond to measurement errors. 
We therefore propose a threshold-based algorithm to identify the \st{real-impossible} \newTxt{unprecedented} events used to test our statistical methods; see Section 2.1 of the Supporting Information for complete details. In short, the method requires that the selected TXx measurements are larger than the 95$^\text{th}$ percentile of all TXx at the station, at least 4$\sigma$ larger than typical extreme temperatures at the gauged location, and have at least one neighboring spatial or temporal measurement that is at least a 1$\sigma$ extreme. We explored various combinations of these subjective thresholds and found that our results were insensitive to these specific choices (see Section 2.1 of the Supporting Information). This approach yields a total of $1,692$ candidate events out of the more than 600,000 measurements from all 7,992 stations. Our algorithm further determines that $n_\text{err} = 147$ are \st{likely} measurement error events (which are discarded from the analysis altogether) leaving $n_\text{real} = 1,545$ real \st{-impossible} \newTxt{unprecedented} events (representing less than $0.3\%$ of all TXx measurements).
The \st{real-impossible} \newTxt{unprecedented} events are withheld from our various statistical analyses and used as out-of-sample test data. The real vs. measurement error events are tallied for each year in Figure~\ref{fig_where_when}(b)-(c); the geographic distribution of where the \st{real-impossible} \newTxt{unprecedented} events occur are shown in Figure~\ref{fig_where_when}(e); Supplemental Table S1 categorizes the number of stations with real \st{-impossible events} and measurement error events. \newTxt{In light of the differing record lengths discussed in Section~\ref{sec:covariates} and shown in Figure~\ref{fig_where_when}(a), we also show the occurrence rates of unprecedented events each year, normalized by the total number of non-missing TXx records; see Figure~\ref{fig_where_when}(d).} Note that \newTxt{both} the number \newTxt{and relative rate} of \st{real-impossible} \newTxt{unprecedented} events appears to be increasing since about 1960, with trends earlier in the record obscured by large spikes in 1934 (108 events), 1936 (342 events), and 1954 (93 events). Furthermore, the \st{real-impossible} \newTxt{unprecedented} events occur across all global land areas that are sampled by the GHCN-D database.

\subsection{Extreme value analysis} \label{sec:eva}

As described in Section~\ref{sec:intro}, our hypothesis is that many extreme temperatures are deemed ``impossible'' largely because of methodological choices. Specifically, the traditional approach ignores (at least) three important sources of information: (1) year-to-year variability in temperature extremes, (2) nearby locations will experience the same types of heatwaves (climatological dependence, over long time scales), and (3) nearby locations will also experience the same heatwave events (weather dependence, over short time scales). 
We briefly describe how we account for each of these sources of information.

First, 
we suppose the TXx measurements in year $t$ at gauged location $\bs$, denoted $Y(\bs,t)$, arise from a Generalized Extreme Value (GEV) distribution whose parameters depend on space- and time-varying covariates. The cumulative distribution function for $Y(\bs,t)$ is
\begin{equation*} \label{gev_fam}
\mathbb{P}(Y(\bs,t) \leq y) = \exp\left\{-\left[ 1 + \xi(\bs,t)\left(\frac{y - \mu(\bs,t)}{\sigma(\bs,t)}\right) \right]^{-1/\xi(\bs,t)} \right\}
\end{equation*}
\cite[][Theorem~3.1.1, page~48]{Coles2001}, defined for $\{ y: 1 + \xi(\bs,t)(y - \mu(\bs,t))/\sigma(\bs,t) > 0 \}$. 
\newTxt{Following, e.g., \cite{Zhang2010influence}, \cite{sillman}, and \cite{risser2024anthropogenic}, we utilize covariates to describe year-to-year changes in different aspects of the extreme value distribution, assuming} 
\st{We further assume}  
\begin{equation} \label{eq:gevMarg}
\begin{array}{rcl}
    \mu(\bs,t) & = & \mu_0(\bs)+\mu_1(\bs)\text{GHG}_t  + {\mu_2(\bs)\text{ELI}_t}+ {\mu_3(\bs)\text{SPEI}(\bs,t)} + \\[0.5ex]
     & & \hskip4ex {\mu_4(\bs)\text{PNA}_t}  + {\mu_5(\bs)\text{NAO}_t} \\[0.5ex]
\log \sigma(\bs,t) & = &  \phi_0(\bs)+{\phi_1(\bs)\text{GHG}_t}  \\[0.5ex]
\xi(\bs,t) & \equiv & \xi(\bs)
\end{array}
\end{equation}
In other words, 
the center of the GEV distribution $\mu(\bs,t)$ is modeled statistically as a linear function of GHG forcing, the ENSO longitude index (ELI), the Standardized Precipitation Evaporation Index (SPEI), the Pacific-North American teleconnection pattern (PNA), and the North Atlantic Oscillation (NAO). The natural logarithm of the year-to-year variability $\log \sigma(\bs,t)$ is modeled statistically as a linear function of GHG forcing.  The shape parameter, $\xi(\bs)$, governs the upper tail behavior and varies across space but is otherwise time-invariant \citep[as is standard practice for heatwaves; see, e.g.,][]{Philip2020}. 
The GEV formalism allows us to quantify extreme heatwaves using three quantities: 
\begin{enumerate}
    \item Data-driven upper bounds: $b(\bs,t) = \mu(\bs,t) - {\sigma(\bs,t)}/{\xi(\bs)}$. 
    This quantity is well-defined when $\xi(\bs)<0$. 
    Note that the estimated upper bound will always be larger than the largest TXx measurement, reiterating the importance of treating the \st{real-impossible} \newTxt{unprecedented} measurements as out-of-sample when fitting GEV distributions.

    \item Risk probabilities, denoted ${p}(\bs, t; u)$, which quantify the likelihood of exceeding a given temperature threshold $u$ at a location $\bs$ and year $t$. Alternatively, the risk probability can be defined as the inverse of the return interval for $u$.

    \item The $\sigma$-event threshold $\tau(\bs) = [{b(\bs,t) - m(\bs,t)}]/{\sigma(\bs,t)} = -{\Gamma(1-\xi(\bs))}/{\xi(\bs)}$, where $\Gamma(\cdot)$ is the gamma function and $m(\bs,t) = \mu(\bs,t) +\sigma(\bs,t)\big[\Gamma(1-\xi(\bs)) -1 \big]/\xi(\bs)$ is the GEV mean. This threshold quantifies how extreme a temperature measurement must be (number of $\sigma$'s) relative to the GEV mean to be considered ``impossible.'' Note that $\tau(\bs)$ is time-invariant because the shape parameter $\xi(\bs)$ is time-invariant.    
\end{enumerate}
\st{Second, we propose a statistical framework to account for the fact that nearby stations will experience similar heatwave climatologies, wherein the GEV coefficients are a linear combination of spatially-coherent basis functions (related to the approach of Cooley et al. (2007).}
\newTxt{Second, we propose a statistical framework to account for the fact that nearby stations will experience similar heatwave climatologies: in other words, the spatially-varying quantities on the right-hand side of Equation~\ref{eq:gevMarg} should be spatially coherent (i.e., spatially dependent). This idea leverages the intuition of Tobler's first law of geography: ``everything is related to everything else, but near things are more related than distant things'' \citep{Tobler1970}. We thus develop an approach related to conditional independence methods \citep[see, e.g.,][]{cooley2007bayesian,Risser2019} wherein the GEV coefficients are a linear combination of spatially-coherent, compactly supported basis functions. The centroid of each basis function corresponds to the center of an equal-area hexagonal global grid (each cell with area approximately 200,000 km$^2$) with nominal spacing of approximately 500 km; $356$ of the cells have at least one station.
}
Our approach mimics that of regional frequency analysis \citep{Hosking1993}, which supposes that a group of (presumably nearby) sites have the same statistics. \newTxt{Unlike regional frequency analysis, however,} our approach does not require one to identify homogeneous regions and instead simply restricts the GEV coefficients to vary smoothly according to geospatial coordinates (longitude, latitude, and elevation). This restriction ensures that nearby stations will have similar climatological properties.

Finally, we account for the fact that nearby stations will experience the same heatwave events (i.e., accounting for weather dependence) using modern techniques from the spatial extremes literature  \st{a deep learning variational autoencoder that incorporates a max-infinitely divisible copula} \citep{zhang2023flexible}. 
\newTxt{This approach accounts for spatial structure in the realized TXx values in a given year using a flexible copula model. The copula can account for spatially- and temporally-varying dependence in extreme events (e.g., heatwaves may have different spatial structures in the tropics versus the midlatitudes, and also the beginning versus the end of the record) while also allowing the dependence to vary as a function of how extreme the measurements are. The copula is the embedded within a variational autoencoder \citep[an unsupervised learning technique; see, e.g.][]{doersch2016tutorial} to enable scalability to the large, global data set that we set out to analyze in this paper.
}

Ultimately, we consider a set of statistical models for estimating upper bound thresholds, starting with the traditional approach with only a single covariate, here GHG forcing (denoted ``M1''), and sequentially increasing the complexity: adding additional physical covariates (denoted ``M2''), accounting for climatological dependence only (denoted ``M3''), and furthermore accounting for weather dependence (denoted ``M4''); see Supplemental Table S2. For each statistical model, all components will be integrated within a Bayesian framework for uncertainty quantification. \newTxt{For each of the statistical models we use proper but non-informative prior distributions for all statistical parameters. The one exception is the GEV shape parameter, for which we use the maximal data information (MDI) reference prior \citep{northrop2016posterior,Zhang2024} which is the optimal ``noninformative'' prior distribution.}
Our expectation is that M4 will perform the best because it captures the most known structure in the data; however, the hierarchy of models allows us to explicitly assess the relative importance of each innovation. 
\newTxt{Note that we could have considered other combinations of methodologies, e.g., accounting for weather and climate dependence but only using the GHG covariate. However, the spatial aspects of M3 and M4 impose regularization on the covariate coefficients \citep[as with statistical learning methods; see, e.g.,][]{gareth2013introduction}, wherein the regression coefficients can be essentially zeroed out in a data-driven manner when they do not improve the fit of the statistical model.}
\st{In addition to the out-of-sample evaluation described in the next section, we furthermore use the Watanabe–Akaike information criterion (WAIC; Watanabe and Opper, 2010; Gelman et al., 2013) to assess the in-sample performance of each statistical model. WAIC scores for each model are shown in Supplemental Table S3.}
\newTxt{More details on all aspects of the statistical methods are provided in Section 2.2 of the Supporting Information.}

\subsection{Upper bound uncertainty and \st{real-impossible} containment of \newTxt{unprecedented events}} \label{sec:ub_uncertainty}

\begin{figure}[!t]
    \centering
    \includegraphics[width=\textwidth]{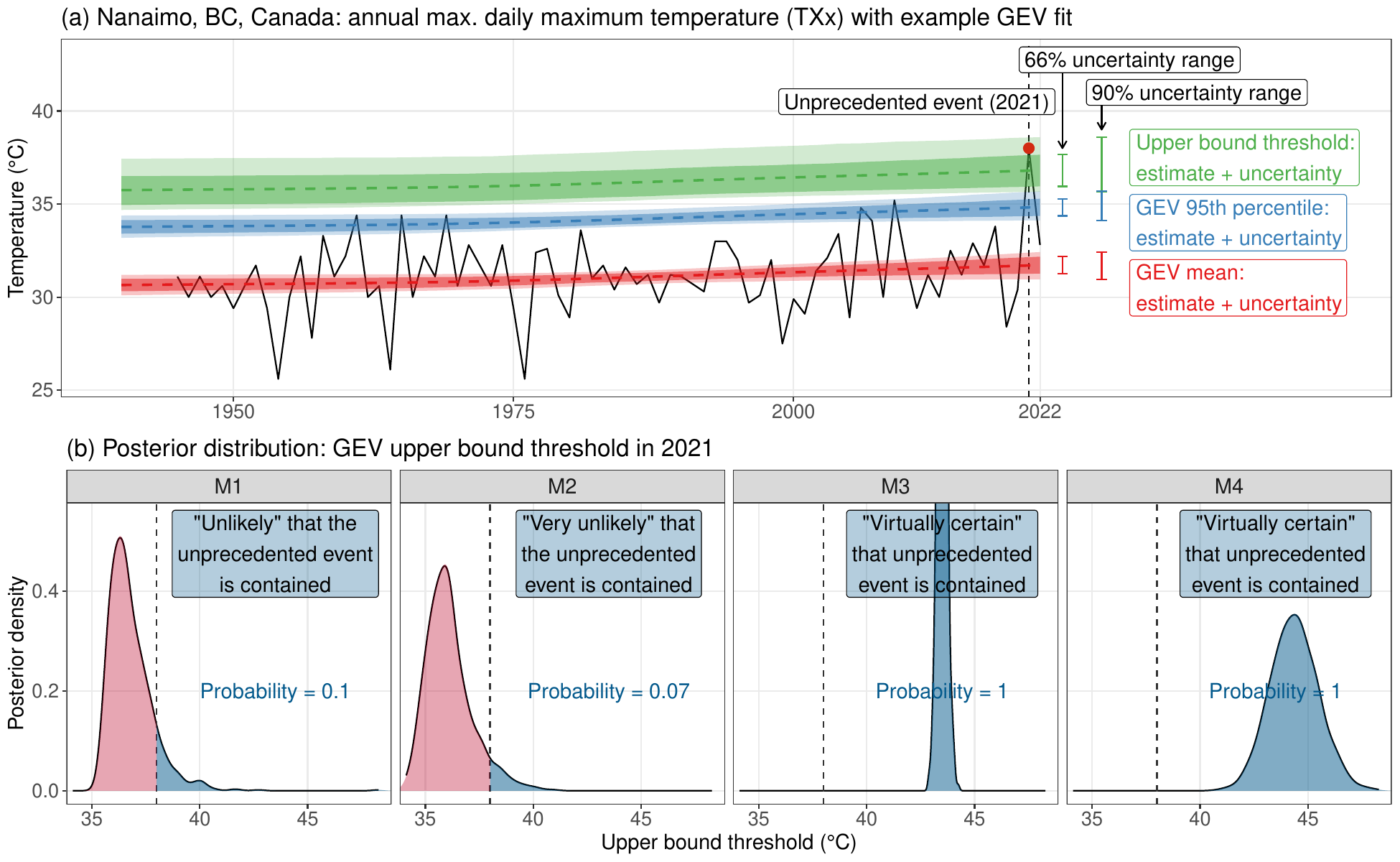}
    \caption{Demonstration of \st{methods} \newTxt{our approach} to account for GEV upper bound uncertainty in determining whether a \st{real-impossible} \newTxt{unprecedented} temperature is contained by a given statistical method. Panel (a) shows TXx records from a gauged location near Nanaimo, BC, Canada that experienced a \st{real-impossible} \newTxt{unprecedented} event of 38.0$^\circ$C in 2021 (red dot), as well as the best estimate (colored dashed lines) and uncertainty intervals (shaded bands) for the mean, 95th percentile, and upper bound threshold of the fitted GEV distribution from M1. Panel (b) shows posterior distributions of the year-2021 upper bound for each statistical method M1-M4 \newTxt{with the corresponding probability that the unprecedented event is contained.} \newTxt{Panel (b) also shows the associated likelihood category \citep{mastrandrea2010guidance}.} \st{and categorizes the required confidence to contain the event (Outcome A; red shading and boxes) as well as the probability that the event is contained (Outcome B; blue shading and boxes). IPCC likelihood categories Mastrandrea et al. (2010) are shown in panel (c).} }
    \label{fig:schematic}
\end{figure}

Recall that our main objective is to assess the extent to which statistical methodology impacts the ``impossiblity'' of each \st{real-impossible} \newTxt{unprecedented} event identified in Section~\ref{sec:selectImposs}. In other words, we want to compare the data-driven upper bound estimates with the temperature thresholds experienced in the \st{real-impossible} \newTxt{unprecedented} events and determine whether or not the upper bound is larger than the actual observed temperatures. Given a finite amount of data (e.g., roughly 120 years of TXx measurements),  estimates of the statistical parameters of a GEV distribution are uncertain, such that each GEV parameter $\mu$, $\sigma$, and $\xi$ are not perfectly known. The implication, then, is that functions of these parameters (e.g., risk probabilities and upper bounds) will also be uncertain. This concept is illustrated in Figure~\ref{fig:schematic}(a), where we show how the TXx measurements imply a statistical fit with uncertainty and how uncertain GEV parameters imply uncertain upper bounds.

Hence, answering the question ``is event $E$ contained by upper bound $U$?'' must account for uncertainty innate to the data-driven upper bound. In a Bayesian sense, all uncertainty is summarized by the \textit{posterior distribution} of the upper bound, which is used to calculate best estimates, uncertainty measures, and credible intervals (the Bayesian version of a confidence interval).
\newTxt{For the purposes of this analysis, we use the posterior distribution to calculate the probability that a unprecedented event is contained by a given upper bound estimate. We subsequently map the estimated probabilities onto the} 
\st{These posterior summaries can be combined with the}
IPCC likelihood scale \citep{mastrandrea2010guidance} 
\st{to describe the probability that a real-impossible event is contained by a given upper bound estimate.}
\newTxt{in order to assign each probability a qualitative confidence label. Note that there is alternate yet complementary way to frame quantifying uncertainty in the upper bounds based on upper confidence limits (see Supporting Information Section 2.3); our results are the same regardless of which perspective is taken.}
\st{The framing of the IPCC likelihood scale (shown in Figure 2c) refers to a specific ``outcome,'' and the different categories refer to the probability of that outcome. For our purposes, then, there are two complementary impossible outcomes that can be assessed:}

\noindent \st{\textbf{Outcome A:} Is the \textit{upper confidence limit} (UCL) of the upper bound is larger than the real-impossible TXx?}

\noindent \st{The UCL refers to the upper endpoint of a one-sided credible interval, and the IPCC likelihood categories come into play via the confidence level of the credible interval. For example, the \textit{likely} UCL would be the 66\% upper limit; the \textit{very likely} UCL would be the 90\% upper limit; etc. The probability of outcome A is then be 1 or 0, since the event either will or will not be contained by a given UCL.}

\noindent \st{\textbf{Outcome B:} What is the likelihood that the observed TXx is  is contained by the GEV upper bound?}

\noindent \st{In a Bayesian sense, functions of statistical parameters (such as the upper bound $b = \mu - \sigma/\xi$) are assumed to have their own probability density function (i.e., the posterior), which allows us to assign a probability between 0 and 1 to Outcome B. This probability can be directly mapped onto the IPCC likelihood categories.}

\noindent \st{Note that these two outcomes are complementary, in the sense that Outcome A is the inverse of Outcome B (up to rounding implied by the categories): an event that is exactly contained by the \textit{very likely} (90\%) upper confidence limit is correspondingly \textit{very unlikely} (i.e., with probability 0.1 or 10\%) to be contained by the upper bound.}

This approach is illustrated in Figure~\ref{fig:schematic}(b), where we plot the posterior distributions of the GEV upper bound from statistical models M1-M4 for the year 2021 at a gauged location near Nanaimo, BC, Canada, that experienced a \st{real-impossible} \newTxt{unprecedented} event of 38.0$^\circ$C during the Pacific Northwest and British Columbia heatwave in the same year (shown in Figure~\ref{fig:schematic}a and denoted by the vertical dashed line in Figure~\ref{fig:schematic}b). 
\st{The left side of each panel of Figure 2(b) quantifies the posterior probability that the event is \textit{not} contained by the upper bound, as well as the \textit{likely} (66\%), \textit{very likely} (90\%), and \textit{virtually certain} (99\%) upper confidence limit. For Outcome A, our IPCC likelihood statement for each model is the lowest confidence level for which the event is contained: for example, with M1, the event is contained by the \textit{very likely} (90\%) UCL but not the \textit{likely} (66\%) UCL. These Outcome A categories are shown in the red boxes in the top left of each panel of Figure 2(b). Conversely,}
The right side of each panel quantifies the posterior probability that the event is contained by the GEV upper bound\newTxt{, which is the area of the posterior density to the right of the unprecedented event threshold}. These probabilities \st{can be} \newTxt{are} mapped onto the likelihood categories to summarize our statistical confidence that the event is contained, which is shown in the blue boxes in the top right of each panel. 
We return to this framework in Sections~\ref{sec:caseStudies} and \ref{sec:res:ub} to tally the evidence for containment using each statistical model across all \st{real-impossible} \newTxt{unprecedented} events.

\subsection{Statistical counterfactuals for event attribution} \label{sec:stat_counterfactual}

Using covariates to describe year-to-year variability and long-term trends in extreme temperatures yields space- and time-varying estimates of the 
risk probabilities, 
such that we can use the fitted statistical models to isolate the human influence on extreme temperatures. This approach follows the ``statistical counterfactual'' methodology proposed in \cite{risser2017attributable} \newTxt{to make Granger-causal \citep{granger1969investigating} attribution statements}: calculate risk probability estimates using a desired combination of GHG forcing, ELI, SPEI, PNA, and NAO. 
Here, we use the GHG forcing time series as a proxy for anthropogenic influence and ELI, SPEI, PNA, and NAO to describe ``natural'' or background conditions associated with extreme temperatures. For each \st{real-impossible} \newTxt{unprecedented} event, we calculate risk probabilities for two climate ``scenarios'' described by specific combinations of the physical covariates: 
\textit{pre-industrial}, with natural conditions (ELI, SPEI, PNA, and NAO) from the year of occurrence and 1901 anthropogenic GHG forcing levels, and \textit{present-day}, with natural conditions from the year of occurrence and 2022 GHG forcing levels.
Both scenarios are counterfactual in the sense that they correspond to climate conditions that did not occur in reality. \st{(except for events that occurred in either 1901 or 2022).}

The risk probabilities are then used 
to quantify the effect of human-induced GHG forcing on the most extreme temperatures. Specifically, we conduct extreme event attribution \citep[EEA;][]{national2016attribution} systematically across all \st{real-impossible} \newTxt{unprecedented} events events to compare the present-day probability of experiencing temperatures at least as extreme as the observed TXx with corresponding pre-industrial probabilities via  
the ``risk ratio'' $RR$ \citep{Paciorek2018}. 
$RR>1$ implies that increases to GHG forcing cause temperatures at least as large as the observed TXx to become more common, while $RR<1$ implies the opposite. Three other cases are possible\newTxt{, all involving cases where the risk probabilities are zero (and hence the event is ``impossible'')}: $RR=\infty$ means that the TXx measurement is impossible \textit{without} climate change (i.e., it has non-zero probability under present-day conditions but zero probability under pre-industrial conditions); $RR=0$ means that the TXx measurement is made impossible by climate change (i.e., it has zero probability under present-day conditions but non-zero probability under pre-industrial conditions); and finally $RR= 0/0$ wherein the risk ratio is mathematically undefined (i.e., the event is impossible in either climate). \st{While traditional approaches to EEA are based on the concept of Pearl causality (Hannart et al., 2015), Philip et al., 2020), 
our statistical counterfactual approach instead utilizes a Granger-causal framework (Granger, 1969) 
using observations only. Granger-causal attribution statements are a somewhat weaker form of causality but are nonetheless useful as they motivate dynamical studies and enhance attribution confidence by using multiple analysis techniques to explore the causes of climate change.}

\section{Results}

\subsection{\newTxt{Checking GEV assumptions and quality of fitted distributions}}

\newTxt{The fundamental challenge in modeling the far upper tail and whether a statistical model can adequately represent the behavior of unobserved extremes beyond the support of the data rests in determining whether the underlying postulates under which the distribution of block maxima may converge to the GEV distribution hold in nature. It is therefore critical to ensure that the GEV assumptions are reasonably satisfied (i.e., that the GEV distribution provides a suitable fit to the data) by the various statistical models proposed in Section~\ref{sec:eva}. One approach for checking the goodness of fit for a statistical distribution to an empirical sample is via a quantile-quantile (Q-Q) plot, which compares the sample quantiles of the data with the corresponding ``theoretical'' quantiles of the fitted distribution. Points falling along the 45$^\circ$ (1-1) line (and within the statistical uncertainty) are evidence of a good fit, i.e., that the sample quantiles are statistically indistinguishable from the theoretical quantiles. Results shown in Section 2.2.5 of the Supporting Information provide strong evidence that the GEV assumptions are satisfied for all four statistical models, and we can be confident in using the fitted distributions to assess the upper tail behavior of extreme daily temperature measurements.}

\begin{table}[!t]
    \caption{In-sample predictive information criteria and out-of-sample log scores for each fitted statistical model. The WAIC, an in-sample metric, is calculated as two times the sum of the log pointwise predictive density (lppd), which summarizes how well the model fits the data, and the effective number of parameters ($p_\text{WAIC}$) to guard against for overfitting \cite{Gelman2013understanding};  smaller scores indicate a better fit. The out-of-sample log pointwise predictive density (lppd-out) summarizes predictive skill for data held out from the process of fitting, here the \st{real-impossible} \newTxt{unprecedented} events; larger lppd indicates a better fit. For each metric, the best model is highlighted with bold text.}
    \begin{center}
\begin{tabular}{ll|ccc|c}
\textbf{Model} & \textbf{\# of GEV} & \textbf{lppd} & \textbf{$p_\text{WAIC}$} & \textbf{WAIC} & \textbf{lppd-out}\\
& \textbf{parameters} & (in-sample) & & & (out-of-sample) \\\hline
M1 & 31,968 & -1,022,407 & 20,090.3 & \textbf{2,084,995} & -10,090.5\\
M2 & 71,928 & -1,003,502 & 49,878.7 & 2,206,761 & -9470.3\\
M3 & 6,052 & -1,117,785 & 9,409.2 & 2,254,388 & -8097.5\\
M4 & 6,052 & -1,491,363 & 214,437.1 & 3,411,600 & \textbf{-6,821.4}\\
\end{tabular}
    \end{center}
    \label{tab_deviance}
\end{table}

\newTxt{Next, we assess the relative quality of the fitted GEV distributions for statistical models M1-M4. We use two quantitative metrics for this evaluation: the Watanabe–Akaike information criterion \citep[WAIC;][]{watanabe2010asymptotic,Gelman2013understanding} to assess the in-sample performance of each statistical model; and the out-of-sample log pointwise predictive density (lppd-out). The WAIC is defined as 
\[
\text{WAIC} = -2\text{lppd} + 2p_\text{WAIC},
\]
where ``lppd'' is the log pointwise predictive density (summarizing the fit of the statistical distribution to the TXx measurements) and $p_\text{WAIC}$ is the ``effective'' number of statistical parameters (used to penalize more flexible statistical models as a way to guard against overfitting). Both quantities are calculated using the fitted posterior distribution and aggregated over all space-time measurements; see \cite{Gelman2013understanding} 
 and \cite{vehtari2017practical} for further details. Smaller WAIC indicates a better model fit. WAIC is commonly used for Bayesian model comparison because it fully captures posterior uncertainty, and it is particular helpful in examining the goodness-of-fit for the entire distribution (including the far upper tail). Therefore, it has been adopted extensively in spatial extremes literature; e.g., \texttt{R-INLA} \citep[]{rue2017bayesian}, skew-$t$ process \citep{hazra2020multivariate}, conditional extremes modeling \citep{simpson2023high}, to name just a few. Since the WAIC evaluates in-sample performance, we also calculate the out-of-sample lppd (i.e., the lppd calculated for the held-out unprecedented events); larger {lppd-out} indicates better model fit.}

\newTxt{WAIC and lppd-out results are given in Table~\ref{tab_deviance}. For statistical models M1-M4,  Table~\ref{tab_deviance} shows the number of statistical parameters in the marginal GEV model, the WAIC and its components (lppd and $p_\text{WAIC}$), and the out-of-sample lppd. The best model is highlighted in bold text for WAIC and lppd-out. First, comparing M1 and M2, note that the lppd for M2 is actually better (larger) than that of M1, which is a direct consequence of M2 having more than twice the number of GEV parameters (71,928 versus 31,968). However, the improvement in the lppd is not enough to offset the increased complexity of the model, such that the WAIC metric prefers M1 over M2. Somewhat surprisingly, the WAIC for M3 and M4 are worse than M1, which indicates that the simplest model provides the best in-sample fit to the data. It is noteworthy, however, that the lppd for M3 is not that much worse than M1 with only about 1/5 as many GEV parameters. When assessing the out-of-sample fit, it is clear that M3 and M4 significantly outperform M1 and M2, with M4 being the best model by a wide margin.}

\newTxt{In summary, the statistical models that account for climate (M3) and weather (M4) dependence yield lppd and WAIC scores that are not too much worse than the baseline model M1 while providing a much better fit to the far upper tail of the fitted GEV distributions. As such, we argue that M3 and M4 have important benefits relative to simpler approaches and provide the best statistical fit for the most extreme temperature events.}

\subsection{Case study events} \label{sec:caseStudies}

\begin{table}[!t]
    \caption{Three \newTxt{unprecedented} temperatures selected as case studies to demonstrate our methodology. The event refers to the annual maximum daily maximum temperature (TXx) recorded at each GHCN gauged measurement site. For reference we also provide the nearest large city to each GHCN site.}
    {\small
\begin{center}    
    \begin{tabular}{lllllr}
       \textbf{Date} & \textbf{GHCN ID} 
       & \textbf{Coordinates} & \textbf{Elevation} & \textbf{Nearest city} & \textbf{Event}    \\ \hline\noalign{\smallskip} 
       29 Jul 2010 & RSM00022602  
       & (30.82$^\circ$E, 63.82$^\circ$N)   & 180.0m & Kostomuksha, RUS & 35.5$^\circ$C\\
       28 Jun 2019 & FRE00106207  
       & (3.96$^\circ$E, 43.58$^\circ$N)    & 2.0m   & Montpellier, FRA & 43.5$^\circ$C\\
       27 Jun 2021 & CA001021830  
       & (124.9$^\circ$W, 49.72$^\circ$N)   & 26.0m  & Nanaimo, CAN & 38.0$^\circ$C\\
    \end{tabular}
\end{center}
}
\label{tab:case_studies}
\end{table}

\begin{figure}[!t]
    \centering
    \includegraphics[width=\textwidth]{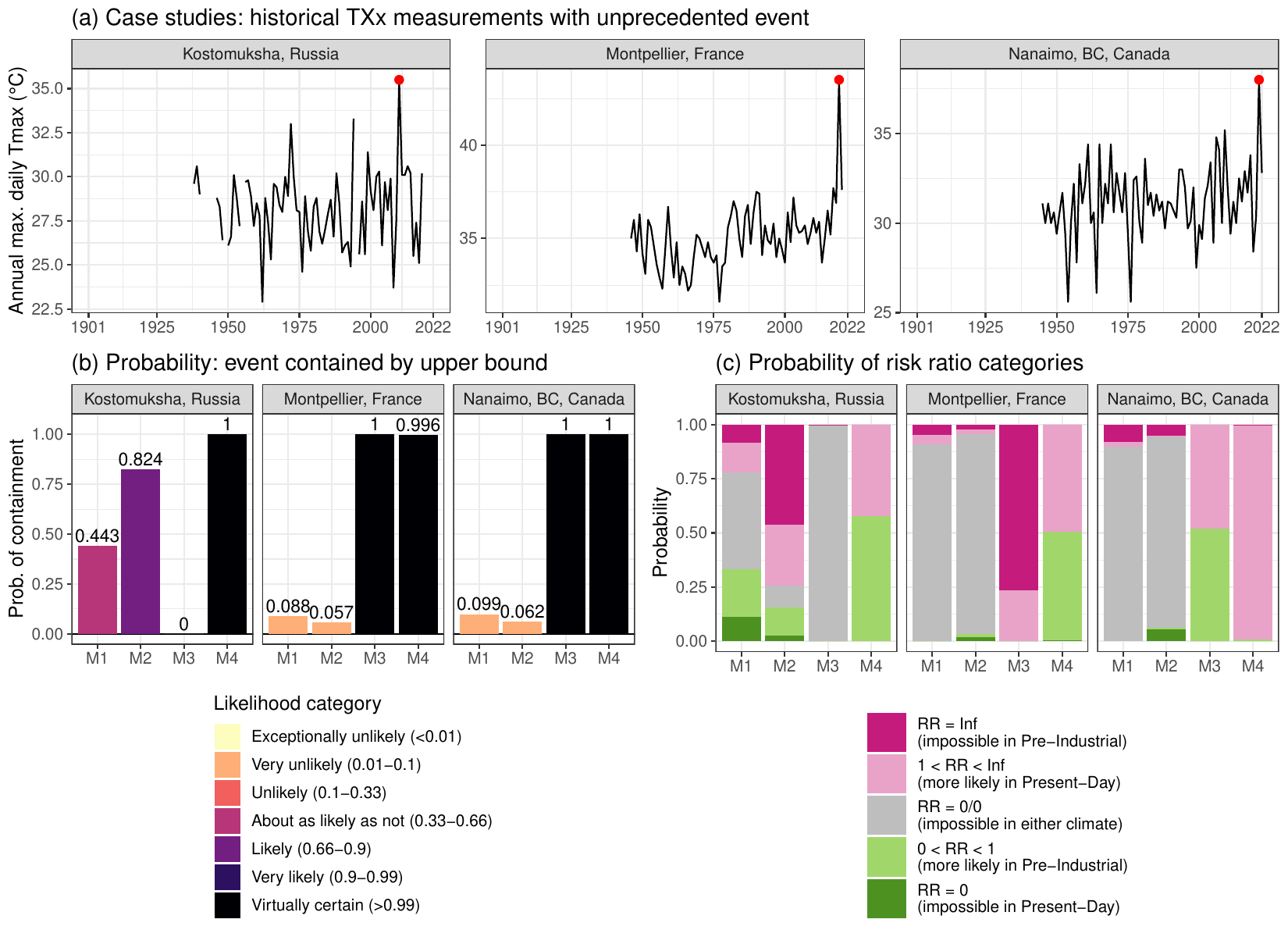}
    \caption{Summary of results for the three \st{real-impossible} \newTxt{unprecedented} events in Table~\ref{tab:case_studies}. Panel (a) shows the historical TXx measurements with the \st{real-impossible} \newTxt{unprecedented} event indicated by a red dot.
    \st{Panel (b) shows the minimum level of certainty required such that the upper confidence limit contains the real-impossible temperature (following Outcome A in Section 2.4).}
    Panel (b) \st{instead} shows the probability that the \st{real-impossible} \newTxt{unprecedented} event is contained by the GEV upper bound\st{ (following Outcome B in Section 2.4). In both panels (b) and (c)}, \newTxt{where} the plotted color maps the \st{minimum uncertainty or} probability of containment onto IPCC likelihood categories. Darker colors indicate better performance.
    Panel (c) tallies the posterior probability that the risk ratio is in each of a set of categories.}
    \label{fig:case_study_results}
\end{figure}

While our final results involve a systematic assessment of extreme events, to demonstrate our methodology we assess three \st{real-impossible} \newTxt{unprecedented} temperatures recorded in the historical record corresponding to the severe heatwave events described in Section~\ref{sec:intro}; see Table~\ref{tab:case_studies}. Historical TXx measurements and the \st{real-impossible} \newTxt{unprecedented} event of interest are shown in Figure~\ref{fig:case_study_results}(a).  In this section, we focus on the event-year upper bound estimates because, as mentioned above, they correspond to the real conditions present during each event.

Following the framework outlined in Section~\ref{sec:ub_uncertainty} and visualized in Figure~\ref{fig:schematic}(b), we tally the 
\st{minimum uncertainty required such that the UCL of the upper bound is larger than the real-impossible TXx and}
the probability that the GEV upper bound contains the \st{real-impossible} \newTxt{unprecedented} TXx in Figure~\ref{fig:case_study_results}(b)\st{ and (c), respectively}. 
\st{The bar height in Figure 3(b) corresponds to the posterior density to the left of the real-impossible TXx (red shading) in Figure 2(b), while} 
The bar height in Figure~\ref{fig:case_study_results}(b) corresponds to the posterior density to the right of the \st{real-impossible} \newTxt{unprecedented} TXx (blue shading) in Figure~\ref{fig:schematic}(b). The plotted colors in Figure~\ref{fig:case_study_results}(b) \st{and (c)} map the \st{minimum confidence required and} containment probabilities\st{, respectively,} onto the IPCC likelihood categories, where darker colors indicate better performance \newTxt{(i.e., higher containment probabilities)} of the statistical method.
Generally speaking, we can see that increasing complexity of the statistical model results in \st{containment with less required certainty and} more probable containment of the \st{real-impossible} \newTxt{unprecedented} TXx. Moving from left (more traditional, M1 and M2) to right (more novel, M3 and M4) across the $x$-axis, the \st{minimum certainty required to contain the} \st{real-impossible} \st{decreases and the} probability of containment for each \newTxt{unprecedented} TXx increases. For all three events, M1 and M2 \st{require relatively high confidence to contain the events and} have low containment probabilities (\textit{unlikely} or \textit{very unlikely}). Notably, M4 contains all three \st{real-impossible} \newTxt{unprecedented} temperatures \newTxt{with a probability of 1} \st{at any confidence level}. M3 has similarly strong performance except for the Russian heatwave, for which it fails to contain the observed TXx. Finally, it is notable that for the French and BC heatwaves, M1 outperforms M2; in other words, including more covariates results in a \textit{smaller} probability of containment. This suggests that a spatial analysis is required when introducing multiple covariates: without the ``borrowing of strength'' enabled by trading space for time \citep[\`a la][]{Hosking1993}, the loss of degrees of freedom caused by adding covariates results in additional statistical noise.  

Why are the upper bounds from statistical methods M3 and M4 so much more likely to contain the \st{real-impossible} \newTxt{unprecedented} events?
To explore this more carefully, Supplemental Figure S14 shows how the GEV location $\mu$, scale $\sigma$, shape $\xi$, and upper bound $b$ depend on methodology for the three case study events as well as three other \st{real-impossible} \newTxt{unprecedented} events from Western Australia (1933), the Central U.S. (1936), and Mexico (1971). There is no systematic way in which the GEV parameters change to increase the upper bounds for M3 and M4 relative to M1 and M2: in some cases it is due to a larger location parameter (e.g., the events in Mexico and Kostomuksha, RUS); in other cases a larger scale (e.g., the events in Australia and Montpellier, FRA); in yet others a larger shape (e.g., the events in Central U.S. and Nanaimo, CAN). In all cases, these changes result in larger upper bound thresholds and larger return levels (return level curves for each of these events are shown in Supplemental Figure S15). These examples illustrate how the GEV upper bound can be influenced by any one of the three statistical parameters of which it is a function of, and how accounting for dependence (weather and climate) can influence the GEV parameters in unpredictable ways.

Finally, equipped with more robust estimates of data-driven upper bounds, we return to the attribution question and assess the extent to which anthropogenic climate change (via increases to GHG forcing) influences the probability of experiencing events that are at least as extreme as the \st{real-impossible} \newTxt{unprecedented} temperatures. Since we have posterior distributions of the GEV parameters, we can similarly obtain posterior distributions of the risk probabilities and hence risk ratios, from which we can calculate posterior probabilities that the risk ratio is in one of five non-overlapping categories: $RR=0/0$, $RR=0$, $0<RR<1$, $1<RR<\infty$, and $RR=\infty$. As with upper bounds, there is a big impact of statistical methodology on risk ratio estimates. In particular, M3 and M4 completely side-step the challenge of the risk ratio being undefined ($RR=0/0$), except for M3 and the Russian heatwave. Furthermore, M4 is able to estimate non-zero risk probabilities for both scenarios, avoiding even the $RR=0$ and $RR=\infty$ outcomes. Focusing in on M4, the best estimates of the risk ratios (posterior median) are 0.87, 0.96, and 6.39 for the three events, respectively, although in each case there is a large probability that $RR>1$ (0.42, 0.495, and 0.992, respectively).

\subsection{Upper bound thresholds for all \st{real-impossible} \newTxt{unprecedented} events} \label{sec:res:ub}

\begin{figure}[!t]
    \centering
    \includegraphics[width=\textwidth]{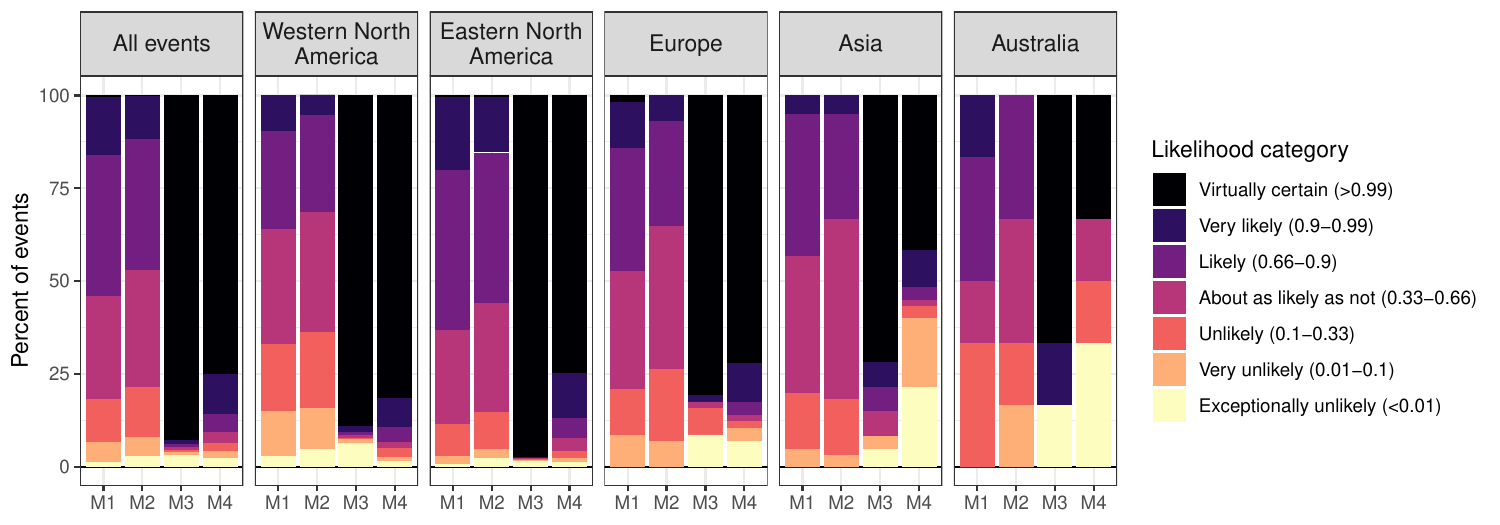}
    \caption{
    The impact of statistical methodology on the probability of containment for \st{real-impossible} \newTxt{unprecedented} temperatures using data-driven GEV upper bound thresholds\st{ (panels a. and b.) and risk ratio best estimates (panel c.)}, aggregated globally and for the continental subregions shown in Supplemental Figure S1.
    \st{Panel (a) tallies the level of certainty required such that the upper confidence limit of the upper bound is larger than the real-impossible TXx (Outcome A in Section 2.4),
    while panel (b) tallies the probability category that the real-impossible TXx is less than the GEV upper bound (Outcome B  in Section 2.4).
    In panels (a) and (b), d}
    Darker colors indicate better performance \newTxt{(i.e., larger containment probabilities)} of the statistical methodology.
    \st{Panel (c) tallies the risk ratio best-estimates (posterior mode) for real-impossible  events for five risk ratio categories.} 
    }
    \label{fig:ub_unc}
\end{figure}

Next, we step back and assess the extent to which all $n_\text{real}=1,541$ \st{real-impossible} \newTxt{unprecedented} temperatures are contained by the upper bound estimates from each statistical model (M1-M4) following the framework proposed in Section~\ref{sec:ub_uncertainty}. 
Figure~\ref{fig:ub_unc} \st{(a)-(b)} tallies the percentage of events overall and in each continental subregion that are contained for each IPCC likelihood category, \newTxt{determined from} 
\st{considering both the level of certainty required such that the upper confidence limit of the upper bound is larger than the real-impossible TXx (Figure~4a, corresponding to Outcome A in Section 2.4) and also the} probability that the \st{real-impossible} \newTxt{unprecedented} TXx is less than the GEV upper bound. 
\st{(Figure 4b, corresponding to Outcome B  in Section 2.4).}
\st{In both panels, d} Darker colors indicate better performance of the statistical methodology, \st{real-impossible} \newTxt{wherein the unprecedented} events \newTxt{are contained with higher probability.} \st{require lower confidence to be contained (panel a) and are more probable to be contained (panel b).}

Three important results emerge from our analysis. 
First, as with the case studies in Section~\ref{sec:caseStudies}, it is clear that increasing the complexity of the statistical model results in better containment of the \st{real-impossible} \newTxt{unprecedented} TXx and therefore fewer ``impossible'' events. Overall, \st{the percentage of events that are contained by (at least) the best estimate of the upper bound goes from 69.1\% for M1 to 92.3\% for M4. Similarly, }
the percentage of events for which the \st{real-impossible} \newTxt{unprecedented} temperatures are \textit{very likely} contained goes from just 16.2\% for M1 to 85.7\% for M4. However, adding complexity does not always improve performance: in many cases the single-station analysis with multiple covariates but no spatial statistics (M2) does not present an improvement over the traditional single-station analysis with only a time trend (M1). For example, the percent of events that are \textit{very likely} contained decreases from M1 (16.2\%) to M2 (11.8\%). It is, however, always true that M3 and M4 outperform M1 and M2.
This suggests that a spatial analysis is required when introducing multiple covariates: without the ``borrowing of strength'' enabled by trading space for time \citep[\`a la][]{Hosking1993}, the loss of degrees of freedom caused by adding covariates results in additional statistical noise. 

Second, \st{the effect of statistical methodology is roughly the same regardless of whether we consider Outcome A or Outcome B. Furthermore,} the effect of statistical methodology is the same regardless of \st{what level of certainty we require (\textit{likely}, \textit{very likely}, etc.) or} what probability threshold is chosen (0.66, 0.9, etc.). For example, the overall percentage of events that are contained depends quite a lot on choice of IPCC confidence category, but for a given category M3 and M4 always represent an improvement over M1 and M2. The largest improvements emerge as one moves from M2 to M3: the percent of events that are (at least) \textit{very likely} contained goes from just 11.8\% for M2 to 93.9\% for M3. This demonstrates that even accounting for climatological dependence only allows us to explain many more events that is otherwise possible. 

Third, our general conclusions hold regardless of what corner of the globe we are considering. For each of the continental subregions shown in Figure~\ref{fig:ub_unc}, as one moves from left (more traditional, M1 and M2) to right (more novel, M3 and M4) across the $x$-axis the colors darken, indicating that more events are contained with less certainty and a higher probability. Furthermore,
\newTxt{in all subregions} the addition of multiple covariates (M2) degrades performance relative to excluding \st{ENSO and SPEI} \newTxt{the drivers of large scale climate variability (as in M1)}. \st{in all subregions.}
The \st{real-impossible} \newTxt{unprecedented} events are the most ``explainable'' in Eastern North America, where 94.4\% of events are contained by (at least) the best estimate under M4. On the other hand, \st{real-impossible} \newTxt{unprecedented} are events are less ``explainable'' in Australia, where just 50\% of events are contained by (at least) the best estimate under M4.

In conclusion, it is clear that whether or not a given historical event is deemed ``impossible'' is largely a function of what statistical methods are used: it is \textit{very likely} that 69.5\% more events are contained by data-driven upper bounds when accounting for both climatological and weather dependence, relative to the traditional approach (16.2\% for M1, compared to 85.7\% for M4).

\subsection{Changes in the likelihood of \st{impossible} \newTxt{unprecedented} events} \label{sec:res:rr}

\begin{figure}[!t]
    \centering
    \includegraphics[width=\textwidth]{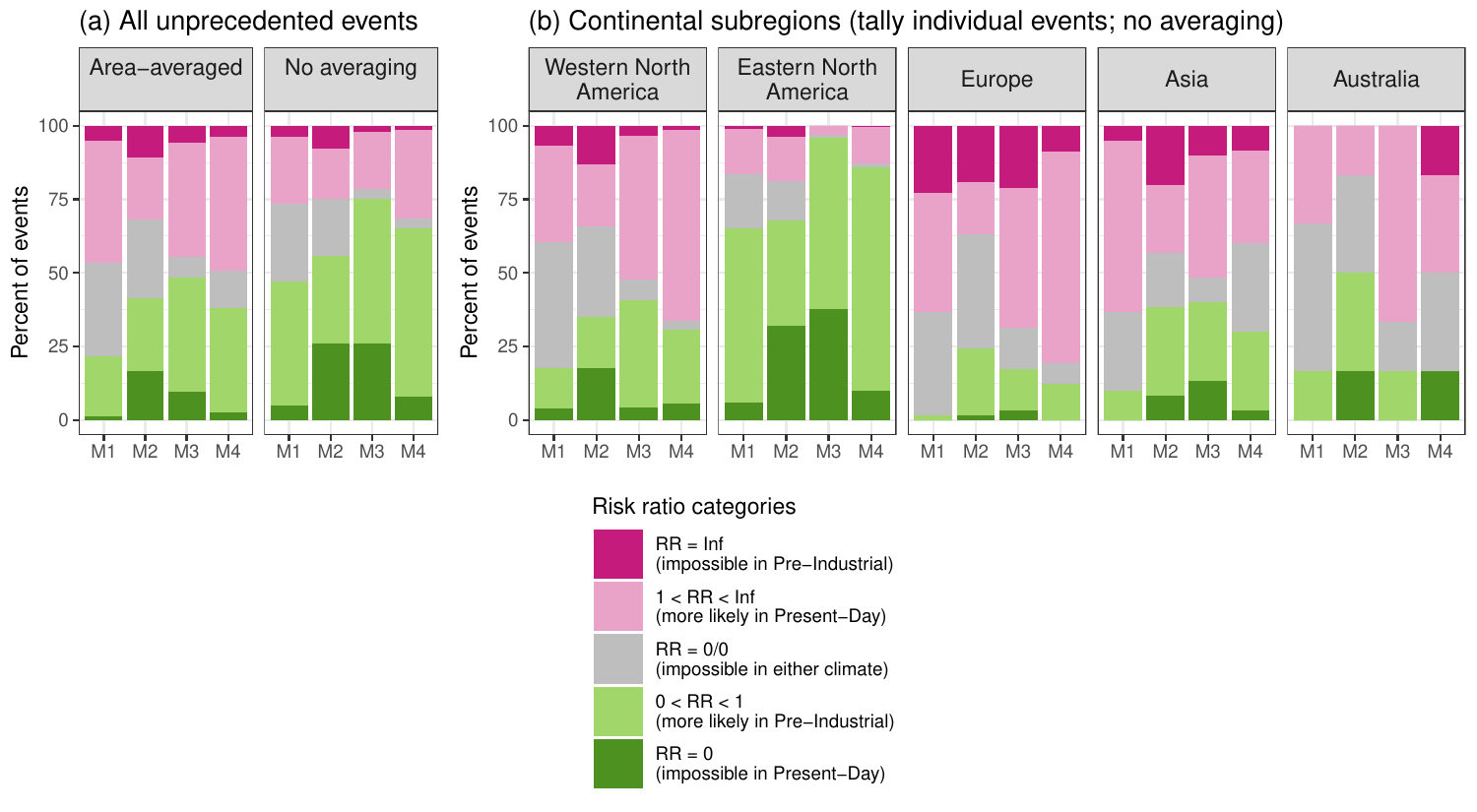}
    \caption{
    \newTxt{The impact of statistical methodology on risk ratio best-estimates (posterior mode) for unprecedented events for five risk ratio categories. Panel (a) summarizes all unprecedented events globally, with and without area-weighted averaging (area averaging accounts for the non-uniform sampling of events). Panel (b) tallies the events separately for each continental subregion shown in Figure~\ref{fig_where_when}. }
    }
    \label{fig:rr}
\end{figure}

Next, for all $n_\text{real}=1,541$ \st{real-impossible} \newTxt{unprecedented} events we calculate best estimates of the risk ratios, here taken to be the posterior mode. The posterior mode is a useful summary when we have ``non-numeric'' outcomes such as $RR=\infty$ and $RR =0/0$; intuitively, the posterior mode is the risk ratio value that is most probable. 
We then categorize the best estimates into the same non-overlapping categories used in Section~\ref{sec:caseStudies}: $RR=0/0$, $RR=0$, $0<RR<1$, $1<RR<\infty$, and $RR=\infty$. 
\newTxt{When considering all of the events globally, we both aggregate the individual events (i.e., with no averaging, ignoring the geographic sampling of events) while also  area-averaging the category probabilities into $200,000$ km$^2$ equal-area hexagonal cells. The latter summary accounts for the non-uniform sampling of the unprecedented events (see panel e. of Figure~\ref{fig_where_when}).}
The percent of risk ratio best estimates in each category is shown in Figure~\ref{fig:rr}(a) for all \st{real-impossible} \newTxt{unprecedented} events \newTxt{(with and without area averaging), while Figure~\ref{fig:rr}(b) shows the events tallied separately} in each continental subregion. 
Similar to our results in Section~\ref{sec:res:ub}, there is a clear effect of methodology on the resulting attribution statements: statistical models M3 and M4 yield uniformly larger numbers of well-defined risk ratios (i.e., not the indeterminate $0/0$). This is a direct consequence of the fact that these methods account for weather and climatological dependence and are hence much more likely to yield non-zero risk probability estimates for pre-industrial and present-day conditions. Also notable is the fact that the dark green $RR=0$ and dark pink $RR=\infty$ categories are generally smaller for models M3 and (especially) M4, again implying that more sophisticated statistical methods allow us to obviate many of the (potentially) hyperbolic statements about climate change making certain events possible or impossible. Similar conclusions hold when considering probabilities of risk ratio categories instead of single-number best estimates; see Supplemental Figure S12.

Interestingly, a large number of risk ratio best-estimates in Figure~\ref{fig:rr} are less than one, i.e., many of the \st{real-impossible} \newTxt{unprecedented} events are more likely in a pre-industrial climate. Risk ratios of less than 1 are due to decreasing trend estimates in GEV statistics, particularly in Eastern North America where the GHCN records are densely sampled. 
\newTxt{A major reason for this has to do with the geographic sampling of the unprecedented events: for M4, the fraction of events with $RR<1$ drops from 65\% without area-averaging to less than 37\% when accounting for the irregular sampling (see Figure~\ref{fig:rr}a.). This sensitivity of attribution results to regional aggregation has been observed elsewhere; see, e.g., \cite{Mindlin2023}. Nonetheless, the surprisingly large fraction of risk ratio estimates less than one} 
\st{this potentially surprising feature} is robust across statistical models and has been verified in different analyses that use independent methods and data sets, see e.g., \cite{Zhang2023upper}. Negative trends in the GEV distribution parameters imply negative trends in the risk probabilities for a fixed threshold, which implies that the \st{real-impossible} \newTxt{unprecedented} events are more probable with the lower 1901 levels of GHG forcing (particularly those towards the beginning of the record).  
Negative trends in temperature extremes in Eastern North America are in part due to the well-documented ``global warming hole'' in this part of the world \citep{Mascioli2017}. The causes of this local cooling are actively debated as to whether it is a manifestation of internal climate variability \citep{Deser2014,Sun2022,Kumar2013} or an externally forced signal \citep{Keil2020,Chemke2020, Qasmi2023}. The contrary appears to be occurring in Europe where very large increases in extreme temperatures have recently been attributed to anthropogenically-driven local circulation changes \citep{Vautard2023}. Hence, a large fraction of European events are deemed impossible without climate change (see Figure~\ref{fig:rr}). 
\newTxt{Note that while a large majority of the Eastern North American unprecedented events occur in the first half of the record (see Supplemental Figure S10), the large fraction of risk ratios of less than one in this region persists whether or not we separately consider events from the full record (as in Figure~\ref{fig:rr}) or from the first half versus second half of the record (see Supplemental Figure S13).}

Ultimately, these regions illustrate both the strength and weakness of Granger causal attribution. In our case, the causal influence of forced local circulation changes are only indirectly represented by our anthropogenic covariate (here, GHG forcing, but equivalently global mean temperature). Granger attribution alerts us to complex changes but does not explain them. On the other hand, with some \textit{a priori} understanding and confidence that the anthropogenic covariates adequately represent the forced changes, Granger causal attribution allows defensible estimation of event probabilities and risk ratios.

In summary, using the best statistical model (M4), the best-estimate of the risk ratio is $\infty$ for 21 of all 1,541 \st{real-impossible} \newTxt{unprecedented} events, meaning that  only about $1.4\%$ of the \st{real-impossible} \newTxt{unprecedented} events are, in fact, ``impossible'' in a pre-industrial climate. However, when we consider only the 314 \st{real-impossible} \newTxt{unprecedented} events from the 21$^\text{st}$ century, when anthropogenic climate change is at its maximum, this percent more than doubles ($3.5\%$ or 11 of the 314 events). Even applying our best statistical methodology, there are still 51 \st{real-impossible} \newTxt{unprecedented} events that have a best estimate of the risk ratio that is undefined, meaning the events are ``impossible'' (i.e., have a probability of zero) in either a pre-industrial or present-day climate. However, this is a significant improvement over the 406 \st{real-impossible} \newTxt{unprecedented} events with $RR=0/0$ under the traditional approach (M1).

\subsection{Relative extremity of the most severe temperatures}


In light of our results on the changing probabilities of unprecedented heatwaves, a final question is: just how extreme may the next record-breaking heatwave be? This is clearly a challenging question from the annual perspective taken in this paper, e.g., the dependence of heat extremes on climate change, large-scale modes of climate variability, and drought conditions -- not to mention the complicated meteorological conditions associated with the most extreme temperatures \citep[see, e.g.,][]{McKinnon2022,Wang2022,Mo2022} which are not considered in our analyses. To obviate aspects of this challenge, we return to the $\sigma$-event threshold, denoted $\tau(\bs)$, which defines a time-invariant relative threshold for the hottest temperatures based on how many ``$\sigma$'s'' that temperature is from a typical extreme (see Section~\ref{sec:eva}). This metric is already utilized in the literature on the most extreme temperatures: for example, \cite{McKinnon2022} found that the 2021 Pacific Northwest heatwave was a 4.5$\sigma$ event and that such events are represented in a large climate model ensemble. 
As a metric, $\tau(\bs)$ is useful from an adaptation perspective: together with $\tau(\bs)$, knowledge of a typical extreme (a proxy for $m(\bs,t)$, the GEV mean) and the year-to-year variability in extreme temperatures (a proxy for $\sigma(\bs,t)$) provides an approximate estimate of how hot temperatures may become. 

\begin{figure}
    \centering
    \includegraphics[width=\textwidth]{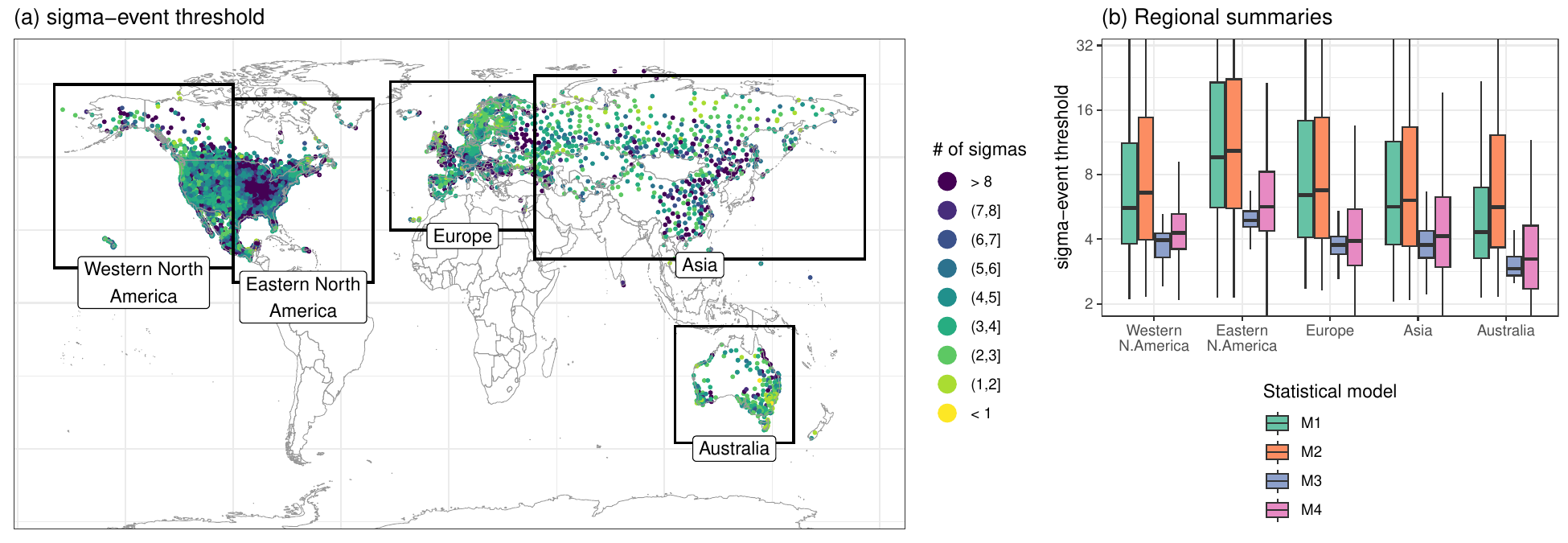}
    \caption{The spatial distribution of sigma-event thresholds $\tau(\bs)$, calculated using statistical model M4, which describes a relative upper bound threshold for how many ``$\sigma$'s'' an extreme temperature might reach (panel a.; plotted color indicates the best-estimate posterior median). Panel (b) shows regional boxplots for the five boxes drawn on the map in panel (a) for all four statistical models.}
    \label{fig:sigma-thresh}
\end{figure}

We now shift our focus from the stations with at least one of the $n_\text{real}=1,541$ \st{real-impossible} \newTxt{unprecedented} events back to all $N=7,992$ gauged locations from the GHCN-D database in order to provide estimates of the relative extremity of the most severe temperatures for the global land regions represented by these stations (note, however, that the outlier and real events are still excluded from the analysis). The spatial distribution of the $\sigma$-event threshold $\tau(\bs)$ calculated using statistical model M4 is shown in Figure~\ref{fig:sigma-thresh}(a), with regional boxplots of $\tau(\bs)$ shown for all statistical models in Figure~\ref{fig:sigma-thresh}(b). Our best estimates from M4 show that most often an ``impossible'' temperature is approximately a 4-5$\sigma$ event. The largest values of $\tau(\bs)$ generally occur inland, e.g., the central United States, eastern Australia, and parts of central Asia, where our method predicts up to 7-8$\sigma$ events. On the other hand, coastal regions have relatively smaller sigma-thresholds, where we estimate that temperatures may only approach 4$\sigma$ events (see, e.g., Western North America, coastal Australia, and northern Europe), reflecting the moderating effect of the oceans on both internal variability and externally-forced changes. The utility of the statistical methods that leverage spatial dependencies (M3 and M4) is further emphasized by the regional summaries in Figure~\ref{fig:sigma-thresh}(b): the single-station analyses of M1 and M2 suggest that events in excess of 20$\sigma$ are possible, which is physically implausible. These results reiterate that single-station analyses involve a large degree of uncertainty and hence introduce statistical noise when estimating upper bound thresholds and probabilities associated with the most extreme temperatures.

\section{Discussion}


In summary, we have clearly demonstrated how the choice of statistical methodology impacts whether the temperatures experienced during the most extreme heatwaves are deemed statistically impossible, as well as the impact on corresponding assessments of the anthropogenic influence on these events. Using the best available statistical tools allows us to anticipate a much larger fraction of the \st{real-impossible} \newTxt{unprecedented} temperatures while minimizing indeterminate risk ratios (i.e., $RR=0/0$). Unfortunately, standard statistical software is only available for the more traditional methods (M1 and M2). In the future, we plan to develop open-source extreme value analysis tools based on M3 and M4  for use by the broader extreme event attribution community.

The other clear message of this paper is that the only way to obtain robust estimates of heatwave probabilities and return intervals is by leveraging the weather and climatological dependence innate to measurements of extreme temperatures. This is particularly important for observational analysis in light of limitations imposed by the relatively short time period from which we have in situ measurements \citep{Zeder2023}. As previously mentioned, while the methods applied here (particularly M4) are cutting-edge even in the statistics literature, the underlying intuition of ``trading space for time'' is a relatively old idea \citep{Hosking1993} that has clear utility for analysis of the most extreme heatwaves.
    
As mentioned in Section~\ref{sec:intro}, a systematic and objective evaluation of extreme heatwaves that considers all global land regions is needed to address outstanding uncertainties regarding low-likelihood, high impact (LLHI) extreme events such as heatwaves \citep{Seneviratne2022_book}. Our results address these uncertainties in two ways: first, the \st{real-impossible} \newTxt{unprecedented} test events we studied occur throughout the 20th century and across the globe, allowing us to avoid the selection bias associated with focusing on more recent events that occur in primarily Western countries. Second, providing non-zero probability estimates for a large majority of the \st{real-impossible} \newTxt{unprecedented} events ensures that we can more robustly study the statistics of the most extreme heatwaves that have occurred in the historical record. Such observational results are critical for evaluating and improving the dynamical models used to develop projections of future climate and the LLHI events that have not yet occurred.

Finally, to accompany this article, we have developed an online graphical user interface (GUI) that allows readers to assess the statistical properties and containment probabilities of an arbitrary \st{real-impossible} \newTxt{unprecedented} event. A link to the GUI is provided in the Open Research section below. The GUI has functionality that allows the user to select a \st{real-impossible} \newTxt{unprecedented} event based on customized longitude-latitude bounds and a given date range. Then, there are several tabs that show all summaries in Sections~\ref{sec:ub_uncertainty} and \ref{sec:caseStudies}: GEV upper bounds and containment probabilities, IPCC likelihood statements regarding containment, return level curves, GEV parameters for the year in which the event occurred, risk probabilities, and posterior probabilities that the risk ratio is in the five categories used in Section~\ref{sec:caseStudies}. While this manuscript presents a summary of all \st{real-impossible} \newTxt{unprecedented} events via aggregation, a reader might be interested in the properties of a specific event beyond those presented in Section~\ref{sec:caseStudies}. Our GUI is designed to provide this specific information. 

\section*{Acknowledgements}

\noindent This research was supported by the Director, Office of Science, Office of Biological and Environmental Research of the U.S. Department of Energy
under the Regional and Global Model Analysis program and the Calibrated and Systematic Characterization, Attribution, and Detection of Extremes (CASCADE) Scientific Focus Area (Contract No. DE-AC02-05CH11231) and used resources of the National Energy Research Scientific Computing Center (NERSC), also supported by the Office of Science of the U.S. Department of Energy, under Contract No. DE-AC02-05CH11231. 

This document was prepared as an account of work sponsored by the United States Government. While this document is believed to contain correct information, neither the United States Government nor any agency thereof, nor the Regents of the University of California, nor any of their employees, makes any warranty, express or implied, or assumes any legal responsibility for the accuracy, completeness, or usefulness of any information, apparatus, product, or process disclosed, or represents that its use would not infringe privately owned rights. Reference herein to any specific commercial product, process, or service by its trade name, trademark, manufacturer, or otherwise, does not necessarily constitute or imply its endorsement, recommendation, or favoring by the United States Government or any agency thereof, or the Regents of the University of California. The views and opinions of authors expressed herein do not necessarily state or reflect those of the United States Government or any agency thereof or the Regents of the University of California.

\section*{Open Research}

\noindent The in situ temperature records supporting this article are based on publicly available measurements from the National Centers for Environmental Information (\url{https://www.ncei.noaa.gov/products/land-based-station/global-historical-climatology-network-daily}).

\noindent A graphical user interface to visualize the results of our analysis is provided at \url{https://mark-risser.shinyapps.io/impossible-temperatures/}.

\bibliography{climate_variability} 

\newpage
\appendix

\begin{large}
\begin{center}
{\Large Supporting Information for:}\\
Data-driven upper bounds and event attribution for unprecedented heatwaves
\end{center}
\end{large}

\section{Data sources} 

\subsection{In situ measurements of daily maximum temperature.}
\begin{figure}[!h]
    \centering
    \includegraphics[width=\textwidth]{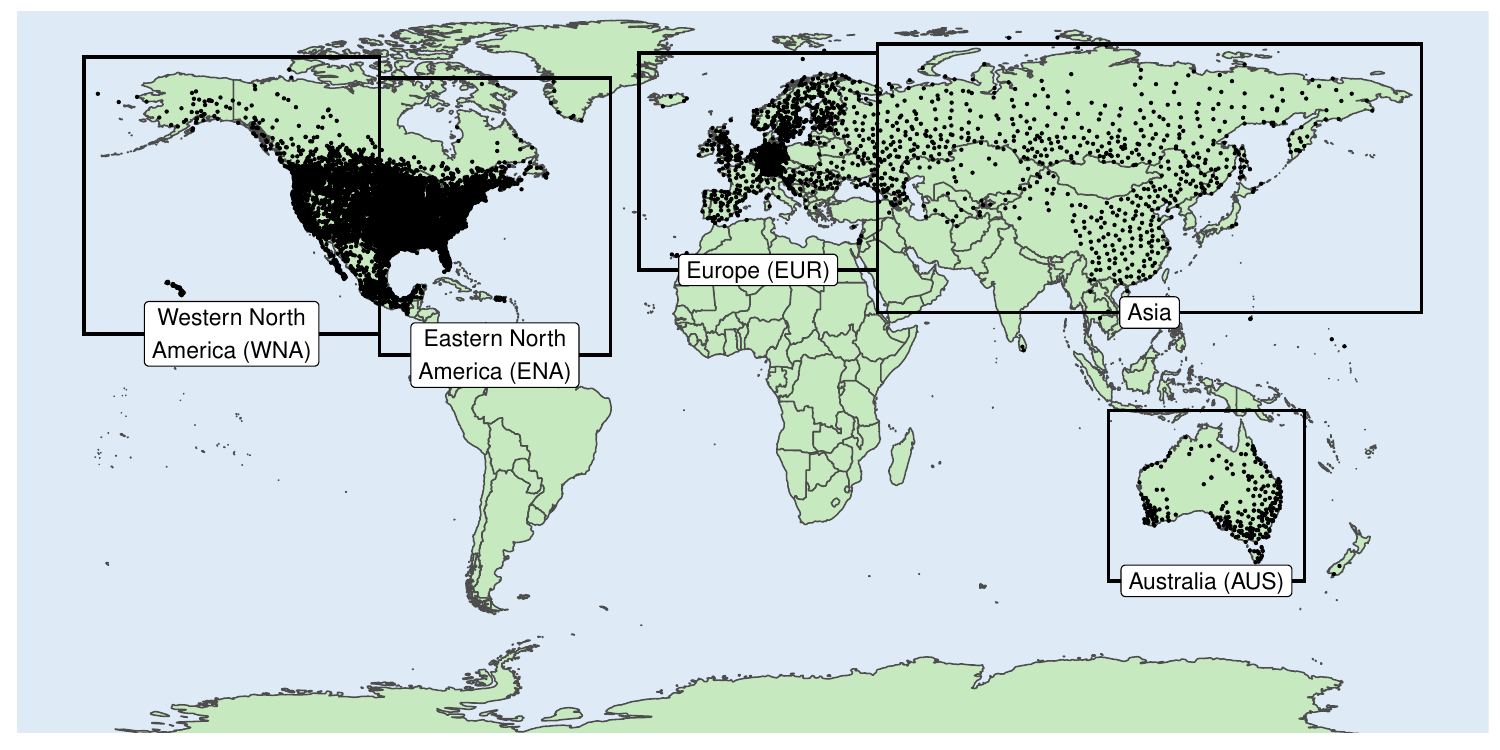}
    \caption{Geographic distribution of the $N=7,992$ gauged locations from the Global Historical Climate Network-Daily database used in our analysis.}
    \label{fig:stations}
\end{figure}

We analyze measurements of daily maximum temperature ($^\circ$C) from the Global Historical Climate Network-Daily (GHCN-D) database \citep{Menne2012} over the historical record, defined as 1901 to 2022. We identify a high-quality set of records based on a minimum threshold of non-missing daily measurements as follows: first, we define annual ``blocks'' as January-December for stations in the Northern Hemisphere and July-June for stations in the Southern Hemisphere. Next, we calculate and store the maximum daily maximum temperature (denoted TXx) in each block-year so long as that block-year had at least 66.7\% non-missing daily measurements. We also require that the TXx occurs in the warm season, i.e., April-September for the Northern Hemisphere and October-March for the Southern Hemisphere. We then select stations that have at least 50 years of non-missing TXx measurements over 1901-2022. Finally, we remove stations that have less than one station within approximately 500km, since this prevents us from leveraging spatially-nearby measurements (this step removes 121 stations from the 8,113 records with at least 50 years of data). Ultimately, this yields $N=7,992$ gauged locations for analysis, denoted $\mathcal{S}$, the geographic distribution of which is shown in Figure~\ref{fig:stations}. Denote the TXx measurements as $\{ Y(\bs,t) \}$ in year $t = 1901, \dots 2022$ for station ${\bf s} \in \mathcal{S}$. Across all stations and years, this yields $n = 612,735$ non-missing TXx measurements for analysis.

\subsection{Physical covariates: year-to-year variability and secular trends} \label{appx_covariates}

\begin{figure}[!h]
    \centering
    \includegraphics[width=\textwidth]{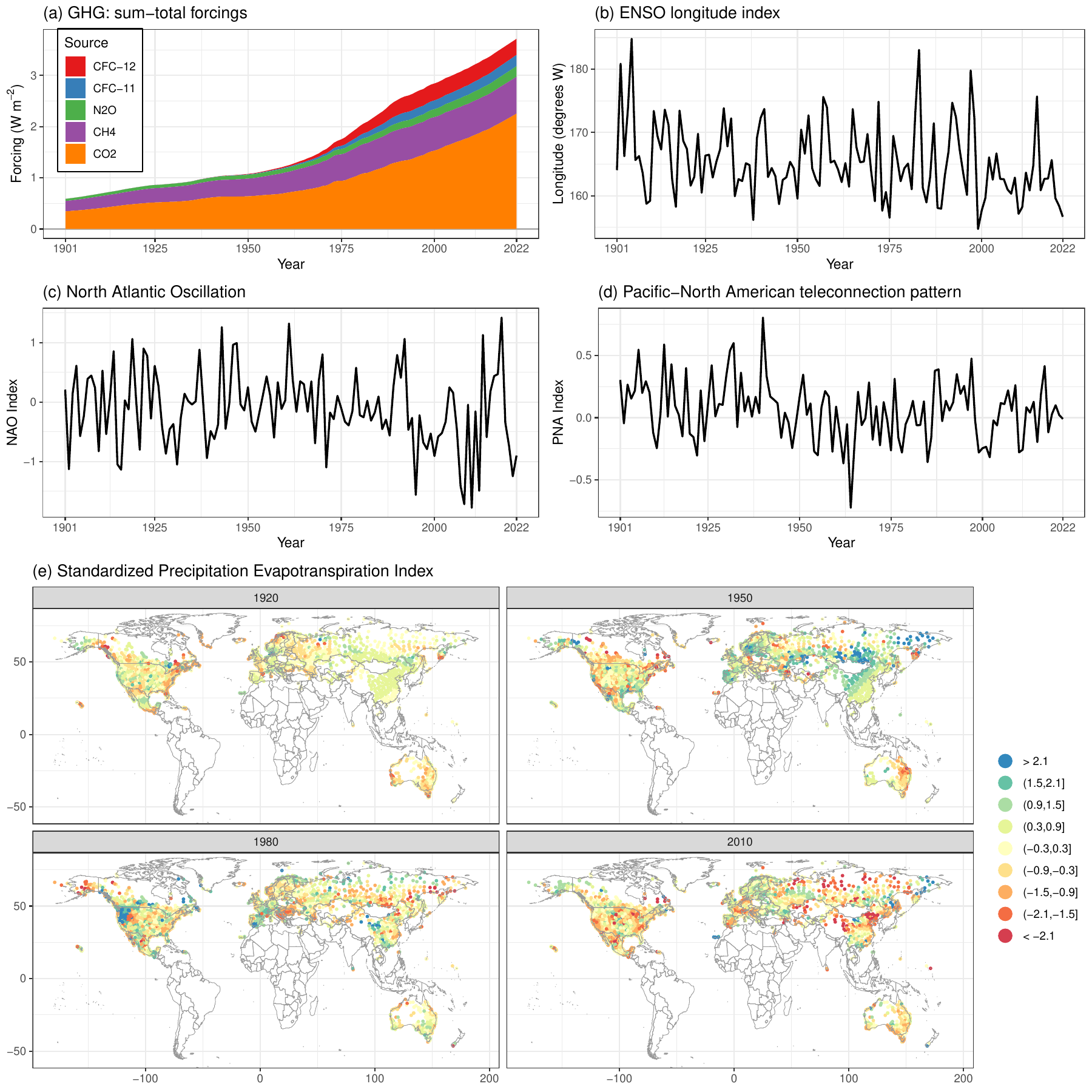}
    \caption{Physical covariates used to describe long-term trends (GHG forcing, panel a.) and year-to-year variability (ENSO longitude index, panel b.; SPEI, panel c.) in the climatology of extreme temperatures; elevation (panel d.) and distance-to-coast (panel e.) used to describe spatial heterogeneity in the statistical parameters that define the climatological distributions. }
    \label{fig:covariates}
\end{figure}

Physical information about the Earth system is a useful tool for describing spatial and spatio-temporal variability in the behavior of weather extremes \citep[see, e.g.,][]{Risser2021quantifying,zhang2023explaining, Zeder2023quantifying}. For our global analysis, we include covariates to describe year-to-year variability and secular trends in the climatology of extreme temperatures. Supplemental Figure \ref{fig:covariates} summarizes these physical covariates.

The covariates chosen for this aspect of the analysis are inspired by the variables used in \cite{zhang2023explaining}. First, dating back to seminal work by \cite{arrhenius1897influence}, it is well-established that anthropogenic greenhouse gas (GHG) emissions drive increases to global mean temperature via radiative forcing of the climate system. Following \cite{Risser2022}, we use the sum-total forcing time series of the five well-mixed greenhouse gases (CO$_2$, CH$_4$, N$_2$O, and the CFC-11 and CFC-12 halocarbons) to describe long-term, human-induced secular trends in the distribution of extreme temperature measurements \citep[see][for further details]{Risser2022}.
Next, the El Ni\~no-Southern Oscillation (ENSO) is a well-known and widely studied mode of natural climate variability that stems from the Pacific oceanic basin. ENSO is a coupled ocean-atmosphere interaction that cycles between positive (El Ni\~no) and negative (La Ni\~na) phases every two to seven years \citep{Philander1985} and is known to have large influence on temperature extremes globally. Among the many metrics used to quantify ENSO, we use the ENSO Longitude Index \citep[ELI;][]{Williams2018diversity} as an additional covariate to represent this influence.  ELI is a sea surface temperature-based index that summarizes the average longitude of deep convection in the Walker Circulation and can account for the different spatial patterns of ENSO. Two other large-scale modes of climate variability have been shown to influence heat extremes globally, namely the Pacific-North American teleconnection pattern (PNA) and the North Atlantic Oscillation (NAO) \citep{Kenyon2008}. Note that \cite{Kenyon2008} actually use the Pacific Decadal Oscillation (PDO); we instead us the PNA for several reasons: first, the PDO index  is highly correlated with ENSO in all seasons; second, \cite{newman2016pacific} show the PDO is not an independent mode but rather an integrator of independent signals at different spatio-temporal scales; finally the PNA is an independent mode of variability (1st EOF of 500mb heights) and has direct physical linkage to extreme weather events.
Lastly, extreme temperatures are highly influenced by  evapotranspirative cooling from the surface soil moisture content and the local vegetation \citep{Domeisen2023}. As such, a measure of drought conditions should be useful for understanding year-to-year changes in extreme temperatures. We use the Standardised Precipitation-Evapotranspiration Index \citep[SPEI;][]{vicente2010multiscalar}, which is a multiscalar drought index that can be calculated from monthly historical temperature and precipitation data -- importantly, this allows us to calculate SPEI for all global land regions back to 1901. We used gridded monthly temperature and precipitation from version 4.07 of the Climate Research Unit (CRU) TS data set \citep{Harris2020} to calculate the SPEI via the \pkg{SPEI} software package for \proglang{R} \citep{R_SPEI}.


\section{Methods}

\subsection{Algorithmic approach for objective selection of real-impossible events} \label{appdx:algorithm}

As described in Section 2.2 of the main text, we require a set of test ``impossible'' events. We utilize the following algorithm to (1) ensure that our candidate events do not suffer from selection bias \citep{Miralles2023}, (2) enable a systematic and objective evaluation of extreme heatwaves that considers all global land regions, and (3) guarantee that the selected events are ``real'' and do not correspond to measurement errors. Our proposed algorithm is as follows:

\begin{enumerate} 
    \item Specify a number of thresholds for quantifying outliers:
    \begin{itemize}
        \item $q_\text{thresh} =$ TXx quantile above which a data point must lie to be considered ``impossible''
        \item $d_\text{min}=$ minimum distance for identifying spatial neighbors
        \item $t_{-}$ and $t_{+}=$ number of days before and after the TXx measurement as temporal neighbors
        \item $z_\text{thresh}=$ number of standard deviations to consider ``impossible''
        \item $z_\text{neighb}=$ number of standard deviations for spatial/temporal neighbors
    \end{itemize}
    \item For each station, calculate thresholded means and standard deviations of the TXx measurements:
    \[
    M({\bf s}) = \frac{1}{|Y(\bs,t) \leq \widehat{F}^{-1}_{\bf s}(q_\text{thresh})|} \sum_{t=1}^T Y(\bs,t) \times I\big(Y(\bs,t) \leq \widehat{F}^{-1}_{\bf s}(q_\text{thresh})\big)
    \]
    \[
    SD({\bf s}) = \sqrt{ \frac{1}{|Y(\bs,t) \leq \widehat{F}^{-1}_{\bf s}(q_\text{thresh})|} \sum_{t=1}^T \Big[Y(\bs,t) \times I\big(Y(\bs,t) \leq \widehat{F}^{-1}_{\bf s}(q_\text{thresh})\big) - M({\bf s}) \Big]^2}
    \]
    (i.e., the empirical mean and standard deviation of all measurements that are less than the empirical $q_\text{thresh}$ quantile).
    \item Identify TXx measurements that are candidate ``impossible'' events:
    \[
    \Big\{ Y(\bs,t): Y(\bs,t) > M({\bf s}) + z_\text{thresh}\times SD({\bf s}) \cap Y(\bs,t) > \widehat{F}^{-1}_{\bf s}(q_\text{thresh}) \Big\}
    \]
    For each of the candidate events, check the following:
    \begin{enumerate}
        \item Check daily Tmax measurements from days in the window $[ t^*-t_{-}, t^*+t_{+} ]$ around the day $t^*$ on which the TXx measurement occurred:
        \begin{enumerate}
            \item If any neighboring daily Tmax measurements are $> M({\bf s}) + z_\text{neighb}\times SD({\bf s})$ 

            Set {\tt indicate\_time\_neighbor} = 1.

            \item Else:

            Set {\tt indicate\_time\_neighbor} = 0 (including cases where all time neighbors have missing measurements).

        \end{enumerate}

        \item Check daily Tmax measurements from neighboring stations (those within $d_\text{min}$ km), denoted $\{{\bf s}_i:i=1,\dots,n_\text{neighb} \}$ for the day $t^*$ on which the TXx measurement occurred:
        \begin{enumerate}
            \item If any neighboring daily Tmax measurements are $> M({\bf s}_i) + z_\text{neighb}\times SD({\bf s}_i)$ (i.e., if extreme relative to that station's extreme statistics)

            Set {\tt indicate\_space\_neighbor} = 1.

            \item Else:

            Set {\tt indicate\_space\_neighbor} = 0 (including cases where all nearest neighbors have missing measurements, AND cases where there are no stations within $d_\text{min}$ km).

        \end{enumerate}

    \end{enumerate}

    \item Categorize as ``real extreme" vs. ``measurement error'' depending on space/time neighbors: 

    \begin{enumerate}
        \item If: {\tt indicate\_space\_neighbor} = 1 $\cup$ {\tt indicate\_time\_neighbor} = 1 $\longrightarrow$ real extreme event.

        \item Else: measurement error $\longrightarrow$ discard measurement.
    \end{enumerate}

\end{enumerate}
Note that all candidate ``impossible'' events are excluded from the extreme value analysis (measurement errors and real). Real extreme events are held out as test data.

In order to assess the sensitivity of our selection of candidate events and subsequent classification as real versus measurement errors, we explored various combination of the subjective thresholds given in \#1 above: $q_\text{thresh} \in \{ 0.8,0.9,0.95,0.97\}$, $d_\text{min}\in \{ 50,100,200\}$, $t_{-}$ and $t_{+}\in \{1,2,3,4 \}$, $z_\text{thresh}\in \{2,3,4 \}$, and $z_\text{neighb}\in \{ 0,1,2\}$.  The total number of candidate events will of course vary drastically, e.g., between $q_\text{thresh} = 0.8$ and $q_\text{thresh} = 0.97$; therefore, we first assess the relative fraction of the candidate events that have {\tt indicate\_space\_neighbor} = 1 or 0 and also {\tt indicate\_time\_neighbor} = 1 or 0. These results are shown in Supplemental Figure~\ref{fig:select_sensitivity}. First, note that for fixed choices of $z_\text{thresh}$ and $z_\text{neighb}$, the relative propotion of labelled points is essentially the same across different values of $d_\text{min}$, $t_{-}$, and $t_{+}$ (i.e., the thumbnail plots within each subpanel are all essentially the same); thus, we set $d_\text{min} = 100$km, $t_{-} = 2$, and $t_{+} = 2$. The number of candidate events and relative partitioning of {\tt indicate\_space\_neighbor} and {\tt indicate\_time\_neighbor} for $d_\text{min} = 100$km, $t_{-} = 2$, and $t_{+} = 2$ is shown in Supplemental Figure~\ref{fig:select_sensitivity2}. Recall that there are $612,735$ non-missing TXx measurements; by definition, ``impossible'' events should be rare, such that there are less than, say, 1\% or $\approx 6,000$ such events. We therefore set $z_\text{thresh} = 4$, since other choices for $z_\text{thresh}$ yield far too many candidate events; for similar reasons we only consider $q_\text{thresh}>0.8$. As a compromise among the remaining options, we select ``middle'' choices of $z_\text{neighb} = 1$ (choosing from 0, 1, and 2) and $q_\text{thresh} = 0.95$ (choosing from $0.9$, $0.95$, and $0.97$). 

Ultimately, these choices yield a total of $1,692$ candidate events, $n_\text{err} = 147$ of which are measurement errors (and discarded from the analysis altogether) and $n_\text{real} = 1,545$ of which are ``real-impossible'' events (these are withheld from the analysis and used as out-of-sample test data). 
Real events and measurement errors are tallied for each station in Table~\ref{tab:impossible_stations} and for each year in Figure 1 in the main text; example real events and measurement errors are shown in Supplemental Figures~\ref{fig:examples_real} and \ref{fig:examples_out}. From Table~\ref{tab:impossible_stations} it is clear that the large majority of stations have neither a real nor measurement error event; otherwise, it is most common for stations that do have a candidate event to have just one over the 1901-2022 period. 
Note that 
the number of real impossible events appears to be increasing since about 1960, with trends earlier in the record obscured by large spikes in 1934 (108 events), 1936 (342 events), and 1954 (93 events).

\subsection{Extreme value analysis} \label{appx:eva}

We now describe three innovations of our methodology relative to the traditional approach that fits single-station GEV distributions with a time trend.

\subsubsection{Nonstationary marginal Generalized Extreme Value analysis} 

First, we propose a flexible framework for statistically modeling year-to-year and long-term changes in the distribution of extreme temperatures via space- and time-varying covariates. Specifically, we suppose the TXx measurements in year $t$ at gauged location $\bs$, denoted $Y(\bs,t)$, arise from a Generalized Extreme Value (GEV) distribution whose parameters vary over space and time as follows:

\begin{equation} 
\begin{array}{rcl}
    \mu(\bs,t) & = & \mu_0(\bs)+\mu_1(\bs)\text{GHG}_t  + {\mu_2(\bs)\text{ELI}_t}+ {\mu_3(\bs)\text{SPEI}(\bs,t)} + \\[0.5ex]
     & & \hskip4ex {\mu_4(\bs)\text{PNA}_t}  + {\mu_5(\bs)\text{NAO}_t} \\[0.5ex]
\log \sigma(\bs,t) & = &  \phi_0(\bs)+{\phi_1(\bs)\text{GHG}_t}  \\[0.5ex]
\xi(\bs,t) & \equiv & \xi(\bs)
\end{array}
\end{equation}
In other words, 
the center of the GEV distribution $\mu(\bs,t)$ is modeled statistically as a linear function of GHG forcing, the ENSO longitude index (ELI), the Standardized Precipitation Evaporation Index (SPEI), the Pacific-North American teleconnection pattern (PNA), and the North Atlantic Oscillation (NAO). The natural logarithm of the year-to-year variability $\log \sigma(\bs,t)$ is modeled statistically as a linear function of GHG forcing.  The shape parameter, $\xi(\bs)$, governs the upper tail behavior and varies across space but is otherwise time-invariant \citep[as is standard practice for heatwaves; see, e.g.,][]{Philip2020}.


The GEV formalism allows us to quantify extreme heatwaves using three quantities, each of which is a function of the GEV parameters $\mu(\bs,t)$, $\sigma(\bs,t)$, and $\xi(\bs)$. First, for a given gauged location $\bs$, when $\xi(\bs)<0$ (as is often the case for extreme temperature), a formal data-driven estimate of the largest possible value that extreme data can reach is

\begin{equation} \label{eg:gev_ub}
b(\bs,t) = \mu(\bs,t) - \frac{\sigma(\bs,t)}{\xi(\bs)}.
\end{equation}
As a data-driven upper bound, it is always the case that $b(\bs,t) > \max_t Y(\bs,t)$; i.e., the estimated upper bound will always be larger than the largest TXx measurement. This reiterates the importance of treating the identifed real-impossible measurements as out-of-sample when fitting GEV distributions.
Second, for a given temperature threshold $u$, we can calculate the so-called ``risk probability,'' or the probability that the TXx measurement in a given year $t$ will be at least as large as $u$:

\begin{equation} \label{eqn:riskProb}
{p}(\bs, t; u) = \left\{ \begin{array}{ll}
1 - \exp\left\{- \left[1- {\xi}(\bs)\left[({\mu}(\bs,t) - u\right]/{\sigma}(\bs,t) \right]^{-1/{\xi}(\bs)} \right\},  & {\xi}(\bs) \neq 0, \\[1ex]
1 - \exp\left\{- \exp\big\{ \left[({\mu}(\bs,t) - u)\right]/{\sigma}(\bs,t) \big\} \right\},  & {\xi}(\bs) = 0
\end{array} \right. 
\end{equation}
These risk probabilities can also be interpreted as the inverse of the recurrence interval for the threshold $u$ and are often used in event attribution statements.
Finally, since $b(\bs,t)$ from Eq.~\ref{eg:gev_ub} defines a threshold for the hottest possible temperatures, we can also calculate the number of $\sigma$'s a temperature measurement can be above a typical extreme such that it becomes ``impossible.'' By a ``typical'' extreme, we use the mean of the GEV distribution which is
\[
m(\bs, t) = \mu(\bs,t) + \frac{\sigma(\bs,t) \Big[ \Gamma\left(1-\xi(\bs)\right)  - 1\Big]}{\xi(\bs)}
\]
(defined when $\xi(\bs)<1$), where $\Gamma(\cdot)$ is the Gamma function. Using this identity, we next define the so-called $\sigma$-event threshold

\begin{equation} \label{eq:sig_thresh}
\tau(\bs) = \frac{b(\bs,t) - m(\bs,t)}{\sigma(\bs,t)} = -\frac{\Gamma(1-\xi(\bs))}{\xi(\bs)}, 
\end{equation}
which is well defined when $\xi(\bs) <0$. 
Note that since the shape parameter $\xi(\bs)$ is time-invariant, the $\sigma$-event threshold is similarly time-invariant.

\subsubsection{Statistical modeling of climatological dependence}

Second, we propose a statistical framework to account for the fact that nearby stations will experience similar heatwave climatologies: in other words, the spatially-varying quantities on the right-hand side of Equation~\ref{eq:gevMarg} should be spatially coherent (i.e., spatially dependent). This idea leverages the intuition of Tobler's first law of geography: ``everything is related to everything else, but near things are more related than distant things'' \citep{Tobler1970}. We thus develop an approach related to conditional independence methods \citep[see, e.g.,][]{cooley2007bayesian,Risser2019} wherein the GEV coefficients are a linear combination of spatially-coherent basis functions. For the location and log scale coefficients $\theta \in \{\mu_0, \mu_1, \mu_2, \mu_3, \phi_0, \phi_1, \phi_2, \phi_3, \xi\}$ in Equation~\ref{eq:gevMarg}, for any $\bs \in \mathcal{S}$,

\begin{equation}\label{eq:spatial_clim}
\theta(\bs) = \sum_{j=1}^{J} x_j(\bs) b_{1,j}^\theta + \sum_{j=1}^{J} x_j(\bs) e(\bs) b_{2,j}^\theta. 
\end{equation}
Here, $\{ b_{k,j}: k = 1, \dots, 2; j = 1, \dots, J\}$ are basis function coefficients to be estimated from the data and $e(\bs)$ is the elevation (meters above sea level) at location $\bs$. 
The $x_j(\bs)$ are compactly supported basis functions

\begin{equation} \label{eq:bumps}
    x_j(\bs) \propto \left\{ \begin{array}{ll}
   \exp\left\{ \left[ 1 - (1-||\mathbf{s}-\mathbf{h}_j||^2/r^2)^{-1} \right] \right\}  & \text{if } ||\mathbf{s}-\mathbf{h}_j||^2 < r  \\
   0  & \text{else}
\end{array} \right. 
\end{equation}
\citep{Noack2017} such that $\sum_j x_j(\bs) = 1$ for all $\bs$, where the $j$th basis function is centered at $\mathbf{h}_j$ and $r$ determines the radius of non-zero support for each basis function. The centroid of each basis function corresponds to the center of an equal-area hexagonal global grid (each cell with area approximately 200,000 km$^2$) with nominal spacing of approximately 500 km; $J=356$ of the cells have at least one station. The hexagonal grids are generated using the \proglang{R} package \pkg{dggridR} \citep{barnes2023dggridR}.
In a Bayesian approach, we additionally propose prior distributions for all the basis function coefficients $\{ b^\theta_{k,j}: k = 1, 2; j = 1, \dots, 356\}$ 
to guard against overfitting, e.g.,

\[
b^\theta_{k,j} \stackrel{\text{iid}}{\sim} N(0, \omega^\theta_k),
\]
where the variances $\omega^\theta_k$ are treated as statistical hyperparameters to be inferred from the data.

In some ways, our approach mimics that of regional frequency analysis \citep{Hosking1993}, which supposes that a group of (presumably nearby) sites arise from the same frequency distribution, i.e., have the same statistics. Our approach does not require one to identify homogeneous regions and instead simply restricts the GEV statistics to vary smoothly according to geospatial coordinates (longitude, latitude, elevation, and distance-to-coast).

\subsubsection{Accounting for the spatial coherence of individual events}


Finally, we propose a statistical framework to account for the fact that nearby stations will experience the same heatwave events, so called ``weather dependence'' or ``data-level dependence.'' This approach M4 integrates both climatological and data-level dependencies using a flexible max-infinitely divisible (max-id) copula approach within a variational autoencoder (VAE) framework to capture spatial extremes. For more details, see \citet{zhang2023flexible}. This approach is similar to the model employed for analyzing the PNW heatwave in \citep{zhang2023explaining}, though their analysis utilized the copula model from \citep{huser2019modeling} with a stationary extremal dependence structure. The nonstationary max-id model extends the work of \citet{reich2012hierarchical} and \citet{bopp2021hierarchical1} by accommodating spatially and temporally varying extremal dependence indices.

Specifically, the copulas of the observed process $\{Y(\bs,t)\}$ and the max-id processes $\{X(\bs,t)\}$ are connected through the following marginal transformation:
 \begin{equation}\label{eqn:marginalTrans}
   F_{Y\vert\gamma(\bs),t}\{Y(\bs,t)\}=F_{X\vert\theta(\bs),t}\{X(\bs,t)\},
\end{equation}
where $F_{Y\vert\gamma(\bs),t}$ is the GEV distribution function with coefficients $\gamma\in\{\mu_0,\mu_1,\mu_2,\mu_3, \mu_4, \mu_5, \sigma_0,\sigma_1,\xi\}$ defined in Eq.~\eqref{eq:gevMarg} and $F_{X\vert\theta(\bs),t}$ is the marginal CDF for max-id process $X(\bs,t)$ parameterized by $\theta(\bs)$. 

The hierarchical construct of the max-id model allows it to be embedded in a VAE \citep{kingma2013auto} and to facilitate the efficient and accurate characterization of extremal dependence across large and complex datasets, addressing limitations of traditional methods that struggle with extreme values and high-dimensional data. The XVAE model is designed to handle nonstationary dependencies over space and time, providing a powerful tool for accurate climate simulations and better uncertainty quantification in extreme events.

\subsubsection{Hierarchy of statistical models}

Ultimately, we consider a set of statistical models for estimating upper bound thresholds, starting with the traditional approach (denoted ``M1'') and sequentially increasing the complexity: adding additional covariates (denoted ``M2''), accounting for climatological dependence only (denoted ``M3''), and also accounting for weather dependence (denoted ``M4''); see Table~\ref{tab:statModels}. Our hypothesis is that M4 will perform the best based on the fact that it captures the most known structure in the data; however, the hierarchy of models allows us to explicitly assess relative importance of utilizing more covariates versus spatial modeling on the marginal distributions versus accounting for the spatial structure of individual heatwave events.

\clearpage
\section{Supplemental tables}

\begin{table}[!h]
    \caption{Tallying the $n_\text{err} = 147$ measurement errors and $n_\text{real} = 1,541$ real ``impossible'' events over stations. }
    \begin{center}
    \begin{tabular}{ccc}
       \textbf{Number of events} & \textbf{Stations with measurement error events}   & \textbf{Stations with real events}    \\ \hline
       0 & 7855 & 6693  \\ 
       1 & 128  & 1084  \\
       2 & 8    & 184   \\
       3 & 1    & 31    \\
    \end{tabular}    
    \end{center}
    \label{tab:impossible_stations}
\end{table}

\begin{table}[!h]
\caption{Summary of the four extreme value analysis statistical models applied to the global TXx measurements from the GHCN records over 1901-2022.}
\begin{center}
{\small 
\begin{tabular}{clcccc}
\hline\noalign{\smallskip}
\multicolumn{2}{c}{\textbf{}} & \textbf{Include} & \textbf{Include ENSO,} & \textbf{Account for}  & \textbf{Account for} \\ 
& & \textbf{GHGs?} & \textbf{SPEI, NAO,} &\textbf{climatology}  & \textbf{weather} \\ 
& &  & \textbf{and PNA?} &\textbf{dependence?}  & \textbf{dependence?} \\ 
\hline\noalign{\smallskip} 
M1 & Pointwise, GHG only (traditional) & $\checkmark$ & & & \\
M2 &  Pointwise, all covariates&$\checkmark$ & $\checkmark$ & & \\
M3 & Conditional independence &$\checkmark$ & $\checkmark$ & $\checkmark$ & \\
M4 & All covariates + max-id copula &$\checkmark$ & $\checkmark$ & $\checkmark$ & $\checkmark$ \\
\noalign{\smallskip}\hline
\end{tabular}
}
\end{center}
\label{tab:statModels}
\end{table}

\begin{table}[!h]
    \caption{In-sample predictive information criteria and out-of-sample log scores for each fitted statistical model. The WAIC, an in-sample metric, is calculated as two times the sum of the log pointwise predictive density (lppd), which summarizes how well the model fits the data, and the effective number of parameters ($p_\text{WAIC}$) to guard against for overfitting \cite{Gelman2013understanding};  smaller scores indicate a better fit. The out-of-sample log pointwise predictive density (lppd-real) summarizes predictive skill for data held out from the process of fitting, here the real-impossible events; larger lppd indicates a better fit. For each metric, the best model is highlighted with bold text.}
    \begin{center}
\begin{tabular}{ll|ccc|c}
\textbf{Model} & \textbf{\# of GEV} & \textbf{lppd} & \textbf{$p_\text{WAIC}$} & \textbf{WAIC} & \textbf{lppd-real}\\
& \textbf{parameters} & (in-sample) & & & (out-of-sample) \\\hline
M1 & 31,968 & -1,022,407 & 20,090.3 & \textbf{2,084,995} & -10,090.5\\
M2 & 71,928 & -1,103,502 & 49,878.7 & 2,306,761 & -9470.3\\
M3 & 6,052 & -1,117,785 & 9,409.2 & 2,254,388 & -8097.5\\
M4 & 6,052 & -1,491,363 & 214,437.1 & 3,411,600 & \textbf{-6,821.4}\\
\end{tabular}
    \end{center}
\end{table}

\clearpage

\section{Supplemental figures}

\begin{figure}[!h]
    \centering
    \includegraphics[width=\textwidth]{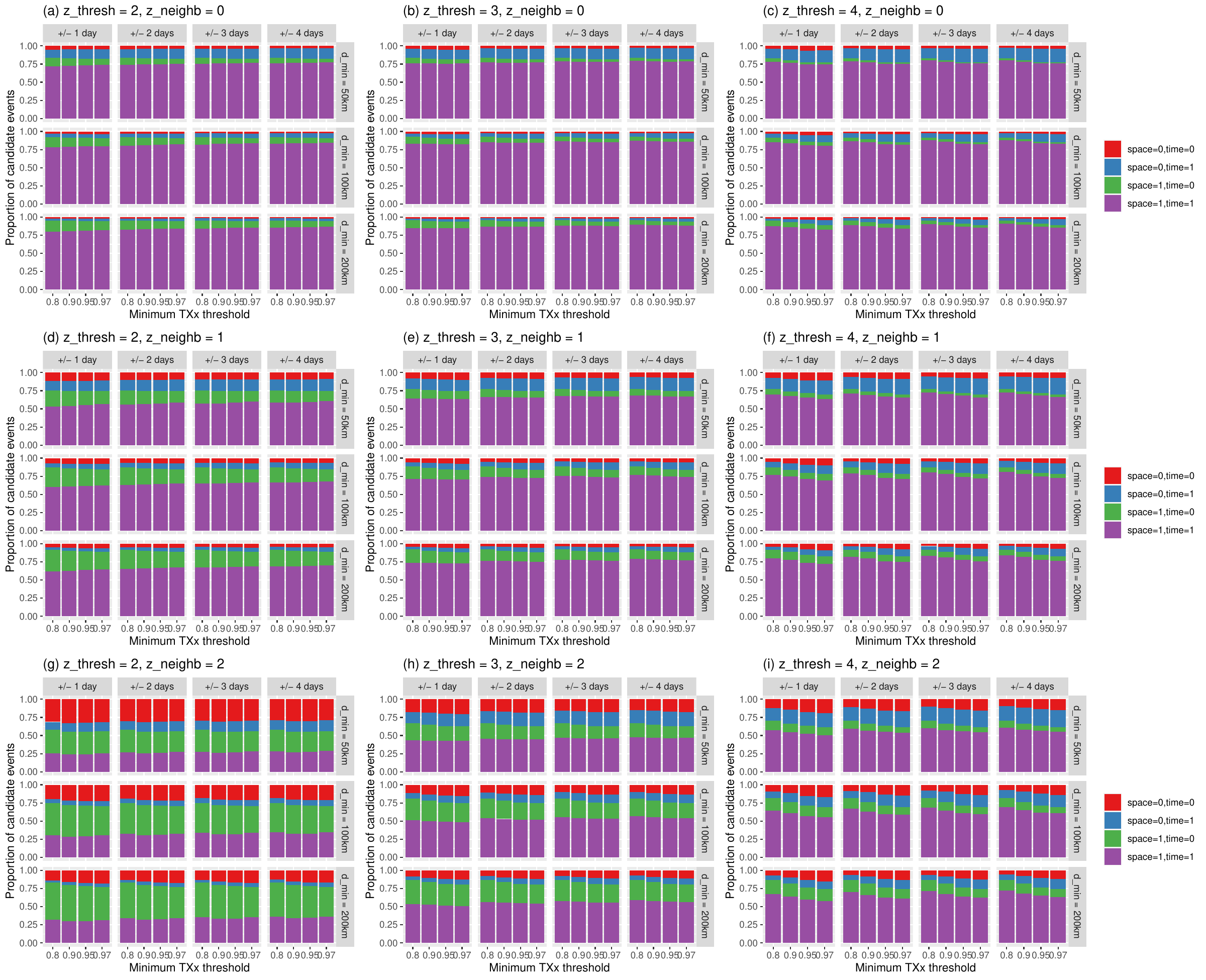}
    \caption{The proportion of events in each category (yes/no spatial or temporal neighbor that is also extreme) across the different thresholds considered for selecting candidate real-impossible events.}
    \label{fig:select_sensitivity}
\end{figure}

\begin{figure}[!h]
    \centering
    \includegraphics[width=\textwidth]{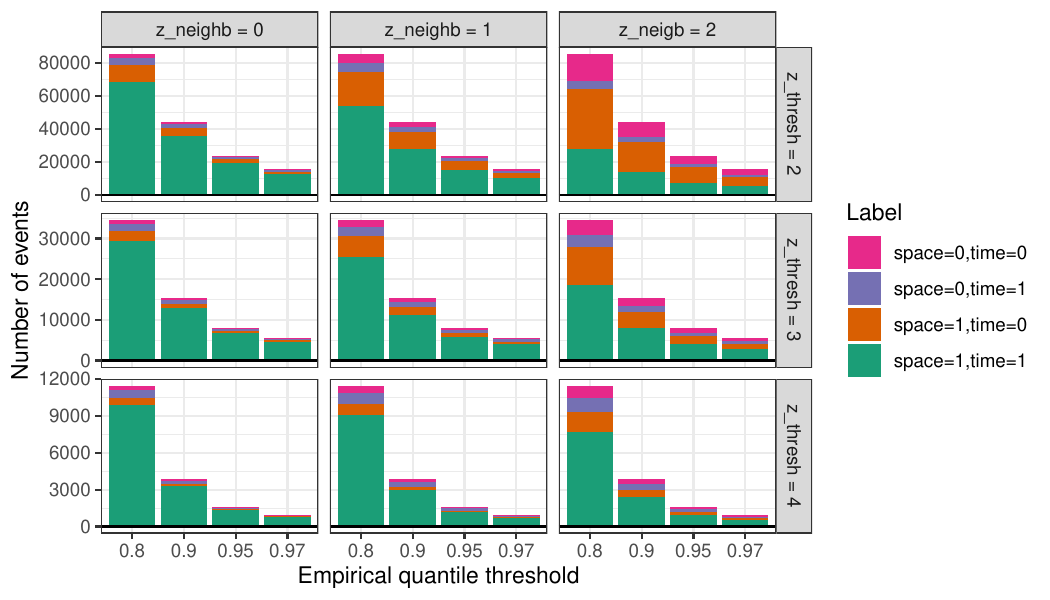}
    \caption{The total number of events in each category (yes/no spatial or temporal neighbor that is also extreme)  across the different thresholds considered for selecting candidate real-impossible events.}
    \label{fig:select_sensitivity2}
\end{figure}

\begin{figure}[!h]
    \centering
    \includegraphics[width=\textwidth]{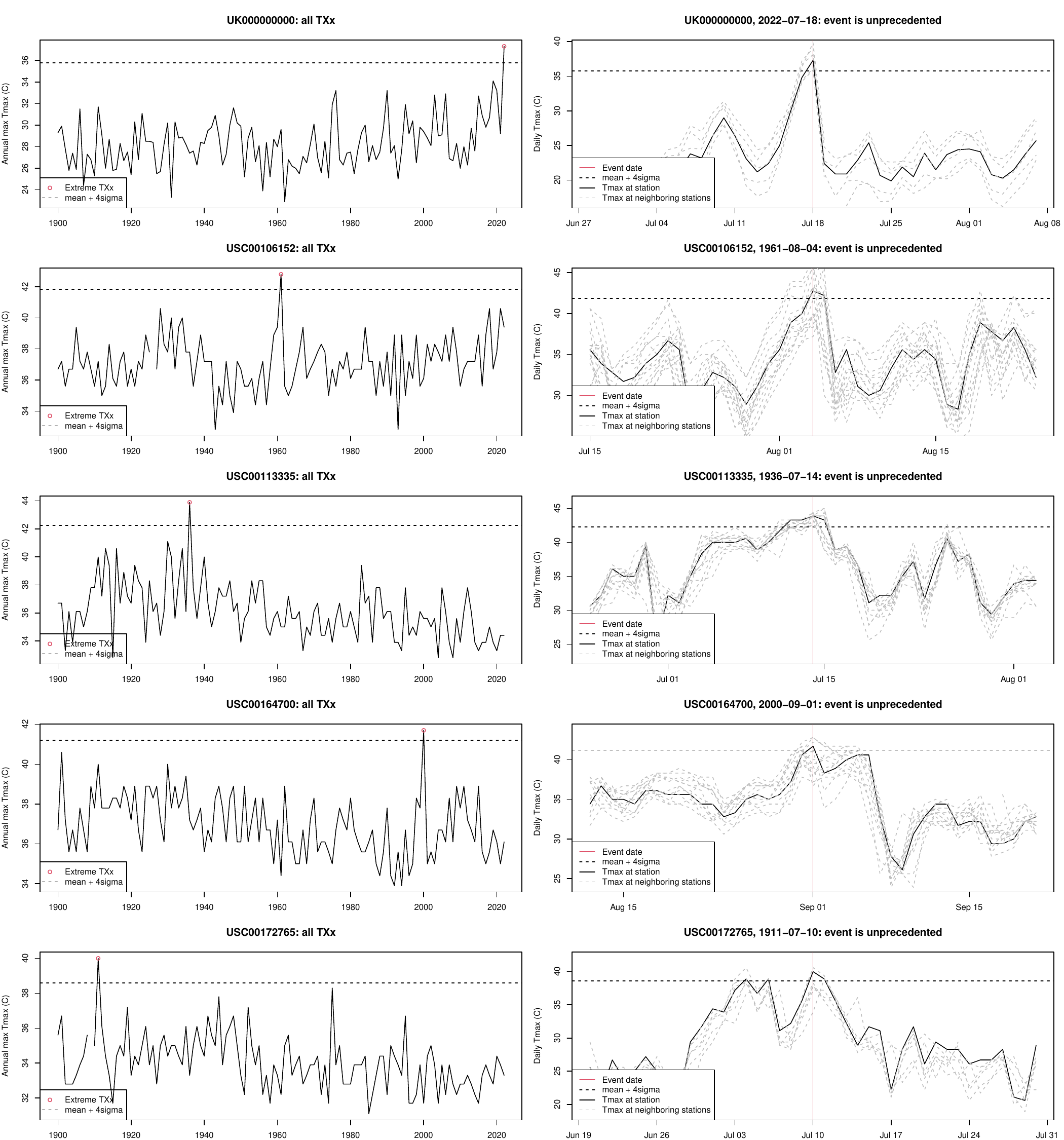}
    \caption{Examples of ``real'' impossible events.}
    \label{fig:examples_real}
\end{figure}

\begin{figure}[!h]
    \centering
    \includegraphics[width=\textwidth]{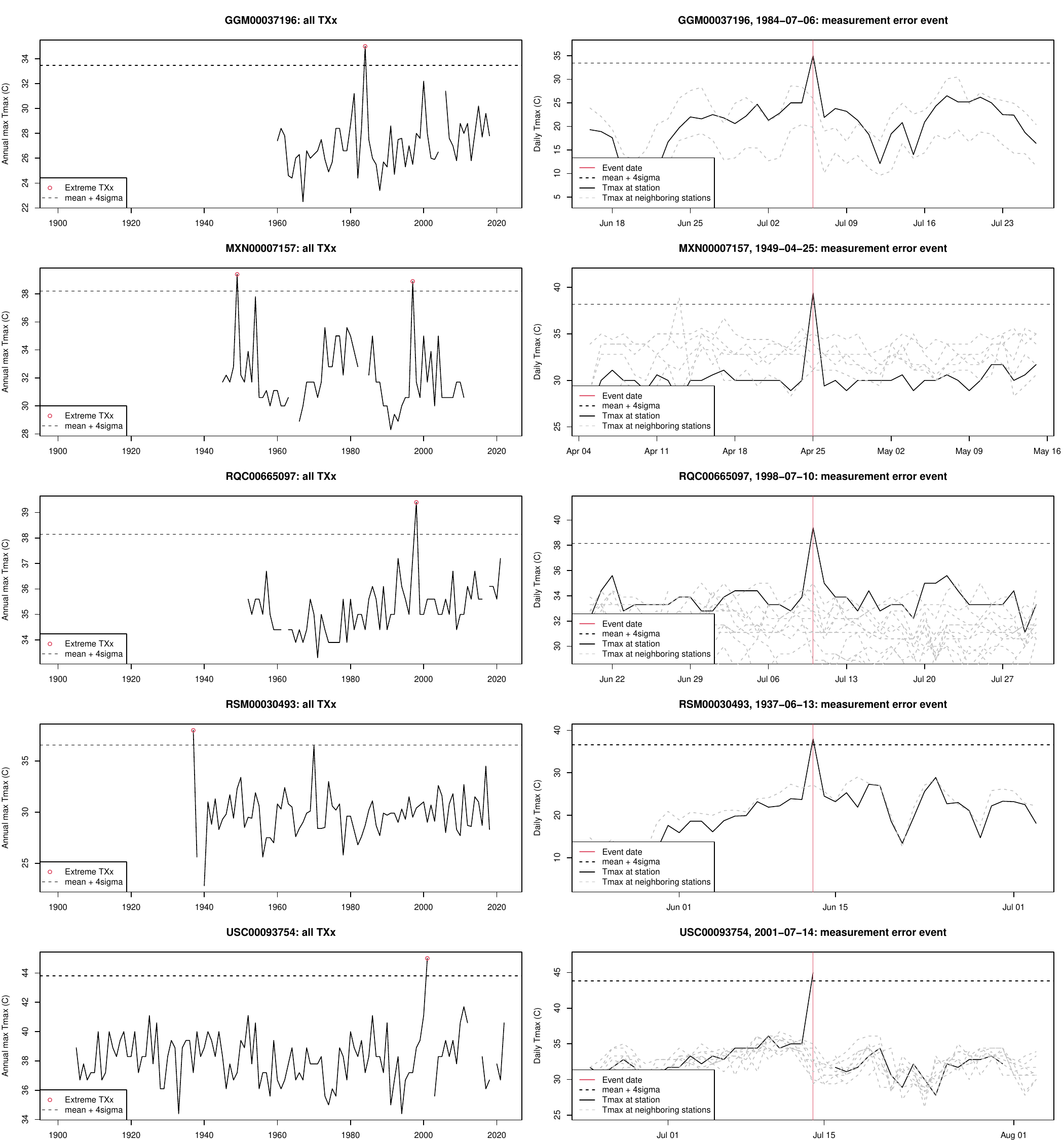}
    \caption{Examples of measurement error events.}
    \label{fig:examples_out}
\end{figure}



\begin{figure}[!h]
    \centering
    \includegraphics[width=\textwidth]{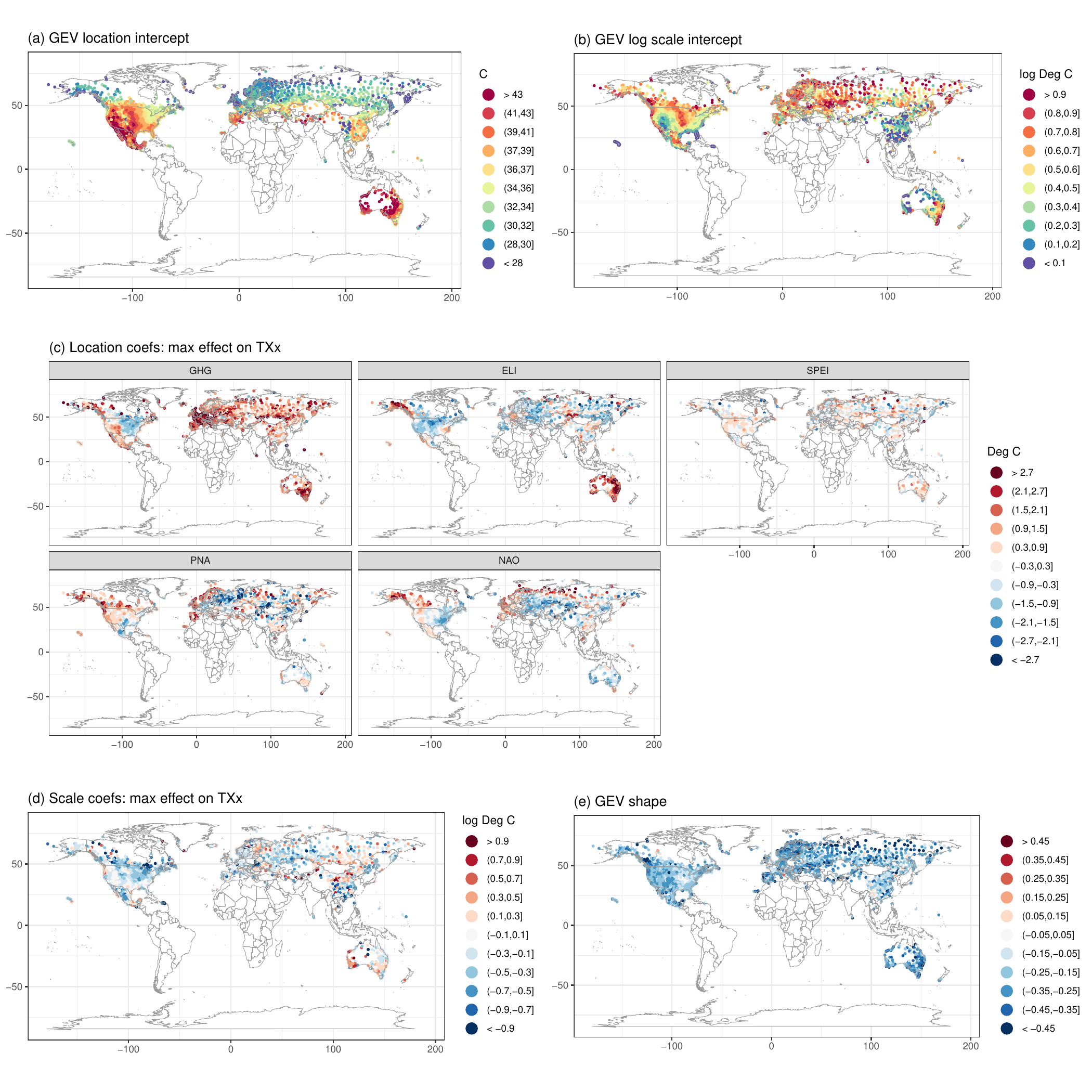}
    \caption{Posterior mean estimates of the spatial distribution of the GEV coefficients obtained from statistical model M4. Panels (b) and (d) show a rescaled version of the coefficients to summarize the largest possible effect each covariate has on the GEV statistics.}
    \label{fig:postmean-coefs}
\end{figure}


\begin{figure}[!t]
\begin{center}
    \includegraphics[width=\textwidth]{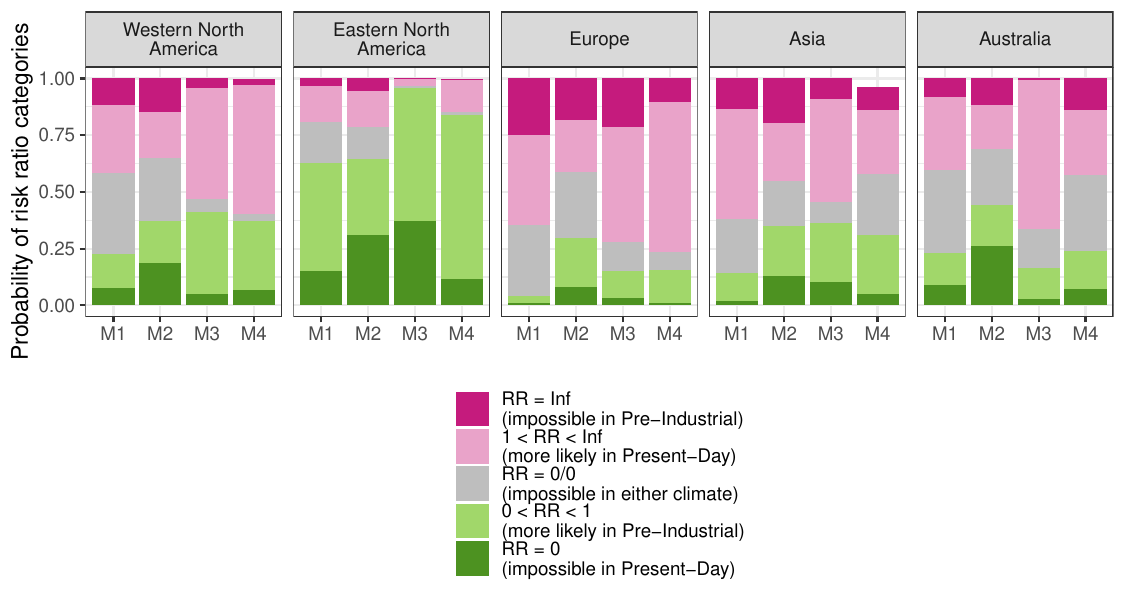}
    \caption{The impact of statistical methodology on posterior probabilities related to risk-ratio-based attribution statements, aggregated over all real-impossible events, 
    both globally and for the continental regions shown in Figure 3 of the main text.
    }
\end{center}
    \label{fig_rr_prob}
\end{figure}


\begin{figure}[!h]
    \centering
    \includegraphics[width=\textwidth]{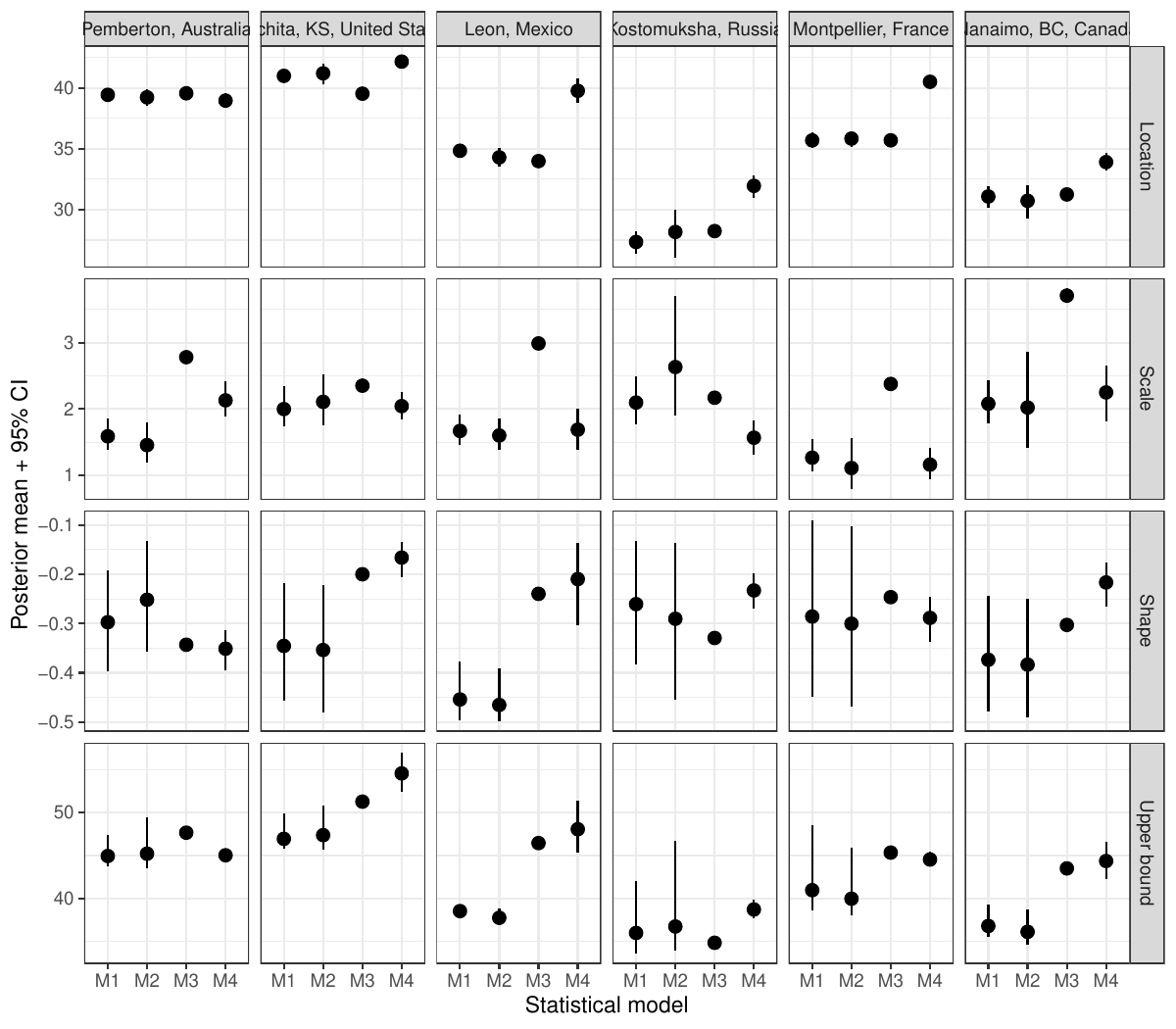}
    \caption{Sensitivity of GEV parameters and upper bounds for six selected real-impossible events.}
    \label{fig:caseStudy_gev_ub}
\end{figure}

\begin{figure}[!h]
    \centering
    \includegraphics[width=\textwidth]{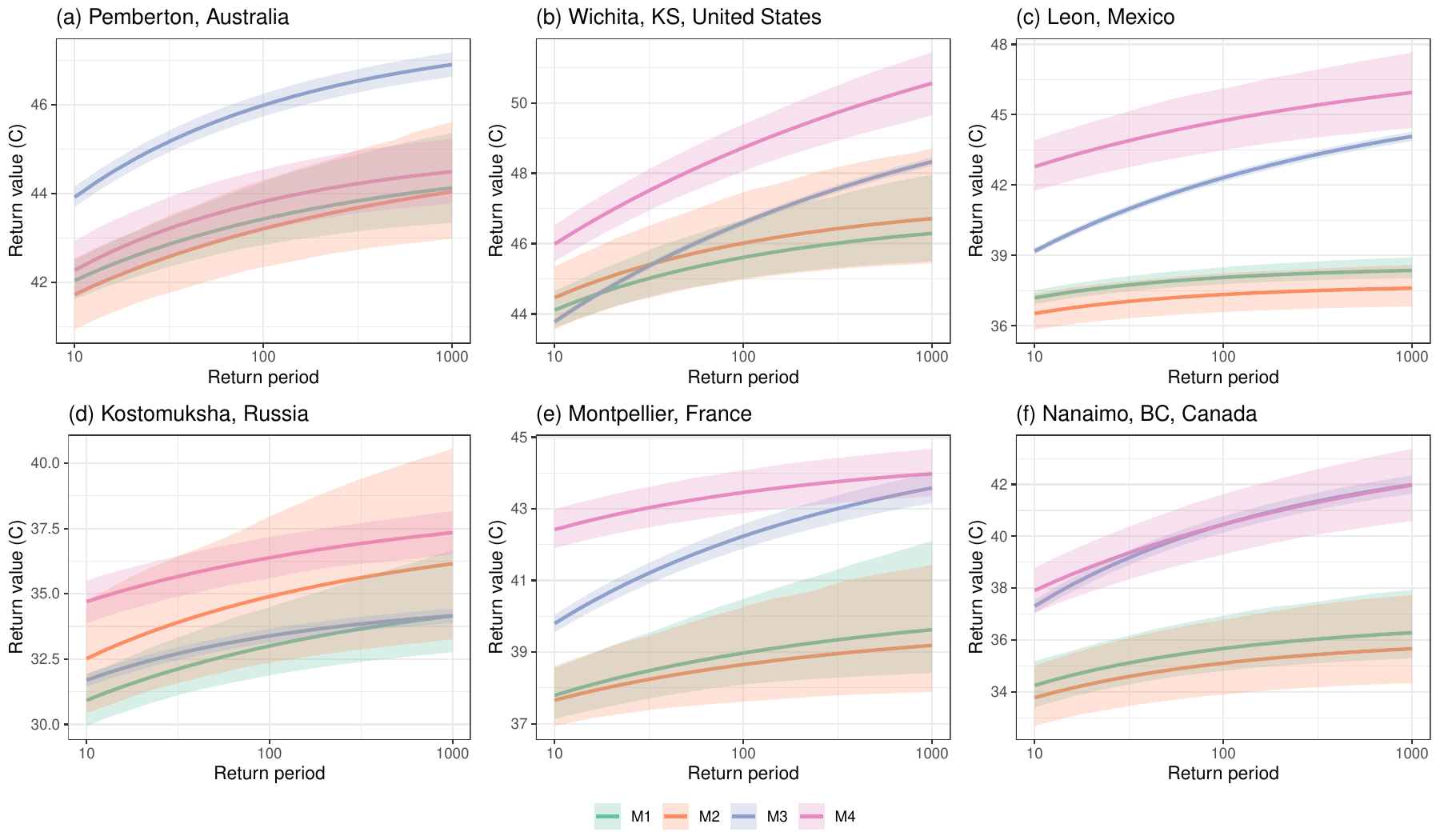}
    \caption{Return levels versus return periods for six selected real-impossible events.}
    \label{fig:caseStudy_rprl}
\end{figure}

\end{document}